\documentclass[prc,twocolumn,tightenlines,showpacs,floatfix]{revtex4}
\usepackage{dcolumn}
\usepackage{bm}
\usepackage{graphicx}
\usepackage{subfig}
\usepackage{hyperref}
\usepackage{xcolor}
\usepackage{soul}
\usepackage{hyperref}

\begin{document}

{
\title{Densities and momentum distributions in $A\leq 12$ nuclei from chiral effective field theory interactions}
\author{M.\ Piarulli$^{1,2}$}
\email{m.piarulli@wustl.edu}
\author{S.\ Pastore$^{1,2}$}
\email{saori@wustl.edu}
\author{\mbox{R.B.\ Wiringa$^3$}}
\email{wiringa@anl.gov}
\author{S.\ Brusilow$^{1}$}
\author{R.\ Lim$^{1}$}
\affiliation{
$^1$\mbox{Department of Physics, Washington University in Saint Louis, Saint Louis, MO 63130, USA}\\
$^2$\mbox{McDonnell Center for the Space Sciences at Washington University in St. Louis, MO 63130, USA}\\
$^3$Physics Division, Argonne National Laboratory, Argonne, IL 60439\\
}

\date{\today}

\begin{abstract}
Current and future electron and neutrino scattering experiments will be greatly aided by a better understanding of the role played by short-range correlations in nuclei.
Two-body physics, including nucleon-nucleon correlations and two-body 
electroweak currents, is required to explain the body of experimental data for both static and dynamical nuclear properties.
In this work, we focus on examining  nucleon-nucleon correlations
from a chiral effective field theory perspective and provide a comprehensive set of new variational Monte Carlo calculations of one- and two-body densities and momentum distributions based on the Norfolk many-body nuclear Hamiltonians for $A\leq 12$ systems.
Online access to detailed tables and figures is available.
\end{abstract}

\maketitle

}

\section {Introduction}
\label{sec:intro}

The coordinate and momentum distributions of nucleons in nuclei are one of the key indicators of short-range correlations (SRCs) in multinucleon systems.  SRCs represent a fascinating aspect of nuclear dynamics; understanding their formation mechanisms and specific characteristics is required to obtain a comprehensive description of nuclei and nucleonic matter.
SRCs tell us much about i) nuclear forces at short distances and how they are generated from quantum chromodynamics; ii) the limitations of mean-field models and how to ameliorate them; iii) the properties of matter at high densities, such as those found in compact stellar objects and in relativistic heavy ion collisions; iv)  the response functions in hadron and lepton scattering from nuclei; v) the origin of the EMC effect, and vi) the sensitivity of neutrinoless double beta decay matrix elements to short-range dynamics.

Since the 1950s, many efforts have been devoted to the study of SRCs and the short-range properties of the nuclear force. It was only recently that experimental and theoretical studies of these phenomena were placed on solid ground, thanks to sophisticated high-energy and large momentum transfer electron and proton scattering experiments~\cite{Tang:2002ww,Piasetzky:2006ai,JeffersonLabHallA:2007lly,Subedi:2008zz,Fomin:2011ng,LabHallA:2014wqo,Hen:2014nza,CLAS:2018xvc,CLAS:2018yvt}, allowing for precision measurements of small cross sections, together with the enormous progress made by many-body theories~\cite{Schiavilla:2006xx,Alvioli:2007zz,Feldmeier:2011qy,Alvioli:2012qa,Rios:2013zqa,CiofidegliAtti:1991mm,Wiringa:2013ala,CiofidegliAtti:2017tnm,CiofidegliAtti:2017xtx}. For instance, experiments involving high-energy, semi-inclusive triple coincidence measurements that successfully probed the isospin composition of nucleon-nucleon ($N\!N$) SRCs in the relative momentum range of 300–600 MeV/c discovered a strong (by a factor of 20) dominance of neutron-proton ($np$) pair SRCs in nuclei when compared with proton–proton ($pp$) and neutron–neutron ($nn$) correlations in both light and heavy nuclei~\cite{Hen:2014nza,LabHallA:2014wqo,CLAS:2018xvc}. This was explained on the basis of the large tensor force in the $N\!N$ interaction at the above-mentioned momentum range. As a result of this finding, it was predicted that the single momentum distributions of the proton and neutron, weighted by their respective fractions, are nearly equal, and that the probability of a proton or neutron being in high momentum $N\!N$ correlation is inversely proportional to their relative fractions in the nucleus. The validity of these predictions were confirmed by results of ab-initio variational Monte Carlo (VMC) calculations of the momentum distributions of light nuclei~\cite{Wiringa:2013ala} and of approximate schemes like cluster expansions~\cite{Alvioli:2012qa,Alvioli:2007zz,Alvioli:2011aa} and correlated basis function theory~\cite{AriasdeSaavedra:2007byz,Bisconti:2007vu,Ryckebusch:2014ann} for medium to heavy nuclei. Moreover, calculations of the momentum distributions of different light nuclei showed high momentum tails that resembled those of the deuteron, demonstrating a universal nature of SRCs~\cite{Wiringa:2013ala,Alvioli:2012qa,Alvioli:2007zz,Alvioli:2011aa,AriasdeSaavedra:2007byz,Bisconti:2007vu,Ryckebusch:2014ann}. \\

An extensive library of VMC one- and two-body densities and momentum distributions for many different light nuclei using the phenomenological Argonne $v_{18}$ (AV18) two-nucleon ($N\!N$)~\cite{Wiringa:1994wb}, and Urbana X (UX) three-nucleon ($3N$) interactions was previously constructed and posted online for the benefit of the nuclear physics community at large~\cite{Wiringa:2013ala}.  Additionally, these calculations have contributed to a novel study of many-body factorization and the position-momentum equivalence of nuclear short-range correlations, using a Generalized Contact Formalism (GCF), which was reported in Nature Physics~\cite{Cruz-Torres:2019fum}. 

In this paper, we provide a comprehensive set of new results of one- and two-body densities and momentum distributions over a wide range of nuclei from $^2$H up to $^{12}$C, using the Norfolk $N\!N$ and $3N$ (NV2+3) forces~\cite{Piarulli:2016vel,Piarulli:2014bda,Baroni:2018fdn,Piarulli:2021ywl,Piarulli:2019cqu}. These results feature new calculations of the pair density as a function of both the pair separation and pair center-of-mass, and calculations of the
two-body momentum distribution coming
from short- and long-range pairs differentiated by a pair separation boundary. The full set of calculations is accessible in graphical and tabular forms online at \url{www.phy.anl.gov/theory/research/QMCresults.html}.

The paper is structured as follows: a brief review of Norfolk interactions is given in Sec.~\ref{sec:norfolk}. In Sec.~\ref{sec:den} we present results for the one- and two-body densities calculated for $^3$H, $^{3,4,8}$He, $^{6,7}$Li, $^9$Be, $^{10}$B, and $^{12}$C. The pair density as a function of both the pair separation and pair center-of-mass is presented for $^{4}$He and $^{12}$C. In Sec.~\ref{sec:mom} the results for the one- and two-body momentum distribution are provided for $^3$H, $^{3,4,8}$He, $^{6,7}$Li, $^9$Be, $^{10}$B, and $^{12}$C. Results for momentum distributions as functions of the relative momentum and center-of-mass momentum without and with pair separation boundary are displayed for $^{4}$He and $^{12}$C.
Additional results are available online.

\section{Norfolk Many-body Interactions}
\label{sec:norfolk}
The Norfolk interactions are obtained from a chiral effective field theory ($\chi$EFT) that uses pions, nucleons and $\Delta$'s as fundamental degrees of freedom, and consists of long-range parts mediated by one- and two-pion exchange, and contact terms specified by unknown low-energy constants (LECs). The LECs entering the $N\!N$ contact interactions are constrained to reproduce $N\!N$ scattering data
from the most recent and up-to-date database collected by the Granada group~\cite{Perez:2013jpa,Perez:2013oba,Perez:2014yla}.
The contact terms are regularized via a Gaussian cutoff function
with $R_S$ as the Gaussian parameter~\cite{Piarulli:2016vel,Piarulli:2014bda,Baroni:2018fdn}. The divergences at high-value of momentum transfer in the pion-range operators are removed via a special radial function characterized by the cutoff $R_L$~\cite{Piarulli:2016vel,Piarulli:2014bda,Baroni:2018fdn}.
There are two classes of NV2 potentials. Class I (II) has been fitted
to data up to $125$ MeV ($200$ MeV). For each class, two combinations
of short- and long-range regulators have been used, namely
($R_S$, $R_L$)=(0.8, 1.2) fm (models NV2-Ia and NV2-IIa) and 
($R_S$, $R_L$)=(0.7, 1.0) fm (models NV2-Ib and NV2-IIb).
Class I (II) fits about 2700 (3700)
data points with a $\chi^2$/datum $\lesssim 1.1$ 
($\lesssim 1.4$)~\cite{Piarulli:2016vel,Piarulli:2014bda}. The short-range component of the $3N$ interactions is parametrized in terms of two LECs, $c_D$ and $c_E$. In the first generation of Norfolk potentials
(NV2+3-Ia/b and NV2+3-IIa/b), these LECs have been determined by simultaneously reproducing the experimental 
trinucleon ground-state energies and $nd$ doublet scattering length~\cite{Piarulli:2017dwd}.
Within the $\chi$EFT framework, $c_D$ is related to the LEC entering
the axial two-body contact current~\cite{Gazit:2008ma,Marcucci:2011jm,Schiavilla:2017}. 
This allows one to adopt a different strategy to constrain $c_D$ and $c_E$.
In particular, in Ref.~\cite{Baroni:2018fdn} they have been constrained
to reproduce the trinucleon binding energies and the empirical value of the Gamow-Teller matrix element in tritium $\beta$ decay. Norfolk models that use this fitting procedure
are designated with a `*' namely, NV2+3-Ia*/b* and NV2+3-IIa*/b*.

These interactions have been recently employed in the VMC and Green's function Monte Carlo (GFMC) approaches~\cite{Carlson:2014vla,Gandolfi:2020pbj} to calculate energies~\cite{Piarulli:2017dwd}, charge radii and electromagnetic form factors~\cite{Gandolfi:2020pbj}, beta-decay transitions~\cite{Baroni:2018fdn,King:2020pza,King:2020wmp}, neutrinoless double beta-decay~\cite{Cirigliano:2019vdj,Cirigliano:2018hja} of light nuclei, beta decay spectra~\cite{King:2022zkz}, muon-capture rates~\cite{King:2021jdb} and with the auxiliary field diffusion Monte Carlo (AFDMC)~\cite{Gandolfi:2020pbj} to study the equation of state of pure neutron matter~\cite{Piarulli:2019pfq,Lovato:2022apd}. 


\section{density distributions}
\label{sec:den}
The one- and two-body densities are evaluated as simple $\delta$-function expectation values given by
\begin{eqnarray}
	\label{eq:rho}
	\rho_{N}(r) &=&\frac{1}{4\pi r^2}\big\langle\Psi\big|\sum_i    \mathcal P_{N_i}\delta(r-|\bf{r}_i-\bf{R}_{\mbox {cm}}|)\big|\Psi\big\rangle\,, \\
	\rho_{N\!N}(r) &=&\frac{1}{4\pi r^2}\big\langle\Psi\big|\sum_{i<j}    \mathcal P_{N_i}P_{N_j}\delta(r-|\bf{r}_i-\bf{r}_j|)\big|\Psi\big\rangle\,, 
\end{eqnarray}
where $\mathcal P_{N_i}$ represents the projector operator onto protons ($+$) or neutrons ($-$)
defined as $\mathcal P_{N_i}=(1\pm\tau_{z_i})/2$, $\bf{r}_i$ is the position of nucleon $i$ and $\bf{R}_{\mbox {cm}}$ is the coordinate of the center of mass.

A detailed survey of one- and two-body densities have been calculated for a variety of nuclei in the range $A=2-12$ using variational Monte Carlo wave functions developed for the AV18+UX and the Norfolk local chiral interactions. The corresponding tables and figures are available online at \url{www.phy.anl.gov/theory/research/density/}, for the one-nucleon densities, and at \url{www.phy.anl.gov/theory/research/density2/}, for the two-nucleon densities.

\subsection{One-body density results}
\label{sec:1den}

\begin{figure*}[!tbp]
 \includegraphics[width=\linewidth]{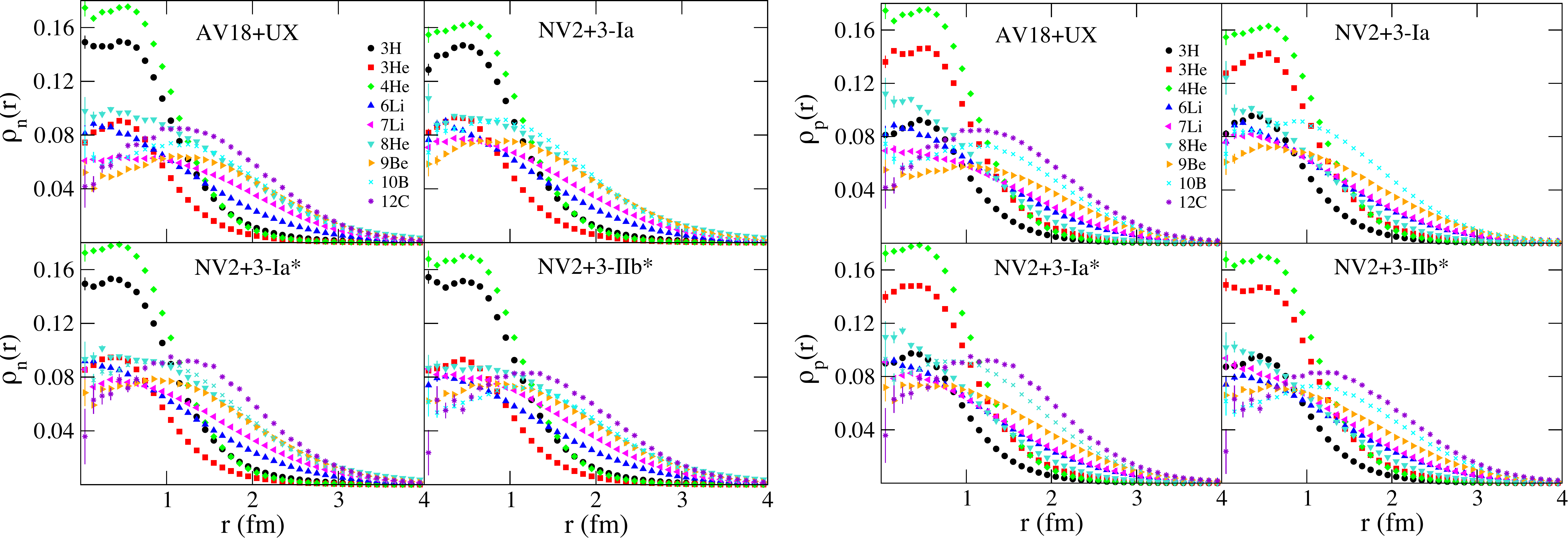}
 \captionsetup{justification=raggedright,singlelinecheck=false}
 \caption{One-body neutron (left panel) and proton (right panel) densities are shown for $^3$H, $^{3,4,8}$He, $^{6,7}$Li, $^9$Be, $^{10}$B, and $^{12}$C using the phenomenological AV18+UX and the local chiral NV2+3-Ia,  NV2+3-Ia*, and NV2+3-IIb* interactions.}  
 \label{1bdensity}
\end{figure*}

In Fig.~\ref{1bdensity} we present the neutron and proton densities calculated for $^3$H, $^{3,4,8}$He, $^{6,7}$Li, $^9$Be, $^{10}$B, and $^{12}$C using the AV18+UX and the NV2+3-Ia,  NV2+3-Ia*, and NV2+3-IIb* local chiral interactions.
Additional densities for $^2$H, $^6$He, $^{8,9}$Li, $^{8,10,12}$Be, $^{11}$B and $^{10,11}$C may be found in the online tables, as well as results for the NV2+3-Ib* and NV2+3-IIa* interactions.
We also give neutron and proton rms radii there.

The VMC wave functions are treated as states of unique isospin $T$.  Thus for $N=Z$ nuclei, proton and neutron densities are the same and only proton densities are given in the online tables.  
However, the wave functions for nuclei with $T>0$ can be different for different isospin projections $T_z$, so mirror nuclei are not isospin symmetric.  This allows the proton-rich nuclei to be slightly more diffuse than neutron-rich nuclei due to their greater repulsive Coulomb interaction.

Spin-up and spin-down densities are also provided in the online tables.  In $J=0$ nuclei, spin-up and spin-down densities are 
identical, but not for $J>0$ nuclei.  
If spin-up and spin-down projections are the same, as in $0^+$ states, we give only totals. The total number of spin-up/down protons and neutrons in $J > 0$ nuclei with $M_J = J$ are reported in Table~\ref{tab:sigma-tau}. Unless otherwise indicated by an error in parentheses, variation among the different interaction models is less than 0.01.
We note that for these nuclei, the subset with an odd number of neutrons has $(n\!\uparrow \!- n\!\downarrow) \approx$ 0.7-0.9, while those with an even number of neutrons have $(n\!\uparrow \!- n\!\downarrow) \approx$ -0.02. Similar results hold for nuclei with odd and even proton numbers.  The sole exception is $^9$Li which has an exceptionally large error bar.

We also note that the s-shell nuclei ($A \leq 4$) exhibit large peaks at small separation, while the p-shell nuclei ($A \geq 6$) are much reduced at small $r$ and more spread out.  This can be attributed to the cluster structure of these light p-shell nuclei, e.g., $\alpha d$ in $^6$Li, $\alpha t$ in $^7$Li, $\alpha \alpha n$ in $^9$Be, and $3 \alpha$ in $^{12}$C.  This puts the center of mass of these nuclei in between clusters and thus reduces the central density. 

\begin{table}[t!]
\captionsetup{justification=raggedright,singlelinecheck=false}
\caption{Total number of spin-up/down protons and neutrons in $J > 0$ nuclei with $M_J = J$ for the local chiral Norfolk NV2+3 interactions.  Variation among the different interactions NV2+3-Ia, -Ia*, -Ib*, -IIa*, and -IIb* is less than 0.01
unless otherwise indicated by an error in parentheses.}
\begin{ruledtabular}
\begin{tabular}{ l d d d d }
Nucleus & \multicolumn{1}{c}{$N_{\uparrow p}$}
        & \multicolumn{1}{c}{$N_{\downarrow p}$}
        & \multicolumn{1}{c}{$N_{\uparrow n}$}
        & \multicolumn{1}{c}{$N_{\downarrow n}$} \\
\hline
$^2$H($1^+$)             & 0.96 & 0.04 & 0.96 & 0.04 \\
$^3$He($\frac{1}{2}^+$)  & 0.98 & 1.02 & 0.94 & 0.06 \\
$^6$Li($1^+$)            & 1.93 & 1.07 & 1.93 & 1.07 \\
$^7$Li($\frac{3}{2}^-$)  & 1.94 & 1.06 & 1.99 & 2.01 \\
$^8$Li($2^+$)            & 1.91 & 1.09 & 2.85(1) & 2.15(1) \\
$^9$Li($\frac{3}{2}^-$)  & 1.91 & 1.09 & 3.12(7) & 2.88(7) \\
$^9$Be($\frac{3}{2}^-$)  & 2.00 & 2.00 & 2.85(2) & 2.15(2) \\
$^{10}$B($3^+$)          & 2.90(1)& 2.10(1)& 2.90(1)& 2.10(1) \\
$^{11}$B($\frac{3}{2}^-$)& 2.87(2)& 2.13(2)& 2.99(1)& 3.01(1)
\label{tab:sigma-tau}
\end{tabular}
\end{ruledtabular}
\end{table}

\subsection{Two-body density results}
\label{sec:2den}

\begin{figure*}[!tbp]
 \includegraphics[width=\linewidth]{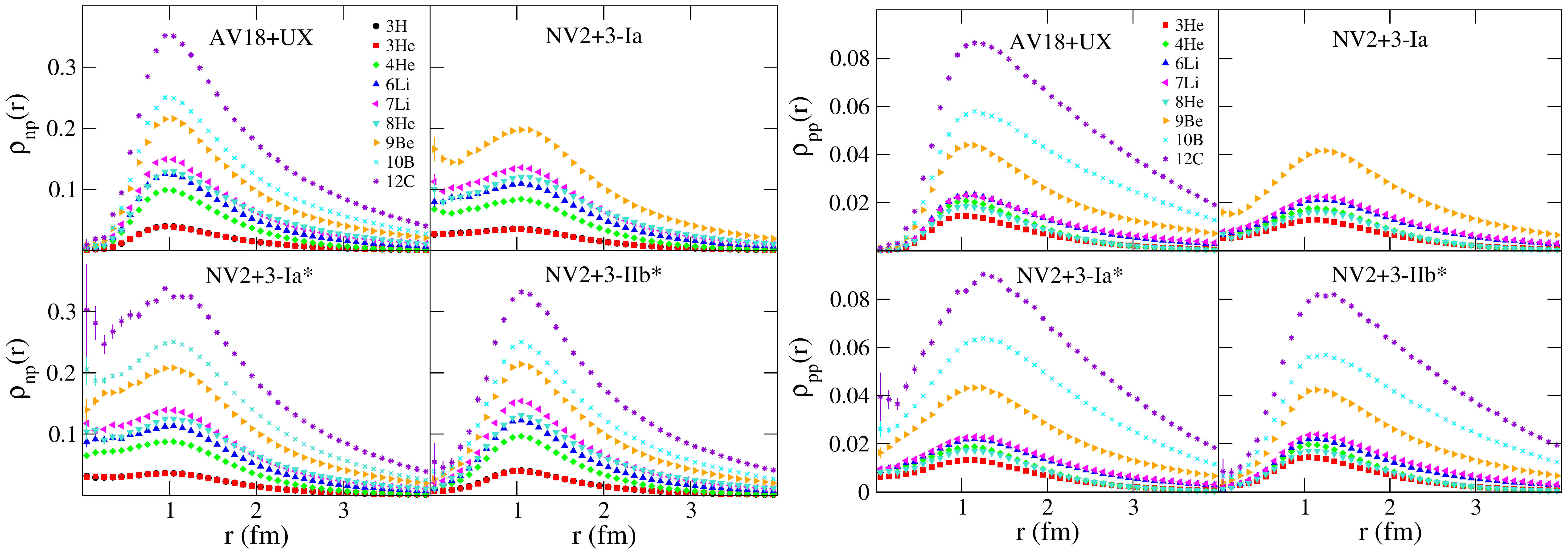}
 \captionsetup{justification=raggedright,singlelinecheck=false}
 \caption{Relative-distance $np$ (left panel) and $pp$ (right panel) pair densities for several
nuclei. Distributions are shown for $^3$H, $^{3,4,8}$He, $^{6,7}$Li, $^9$Be, $^{10}$B, and $^{12}$C using the phenomenological AV18+UX and the local chiral NV2+3-Ia,  NV2+3-Ia*, and NV2+3-IIb* interactions.}  \label{2bdensitiesnoscaled}
\end{figure*}

In Fig.~\ref{2bdensitiesnoscaled}, we present the relative-distance pair densities, with neutron-proton ($np$) in the left panel and proton-proton ($pp$) in the right panel, for $^3$H, $^{3,4,8}$He, $^{6,7}$Li, $^9$Be, $^{10}$B, and $^{12}$C using the phenomenological AV18+UX and the local chiral NV2+3-Ia, NV2+3-Ia*, and NV2+3-IIb* interactions. The online tables contain additional results for the NV2+3-Ib* and NV2+3-IIa* interactions.

\begin{figure*}[!tbp]
 \includegraphics[width=\linewidth]{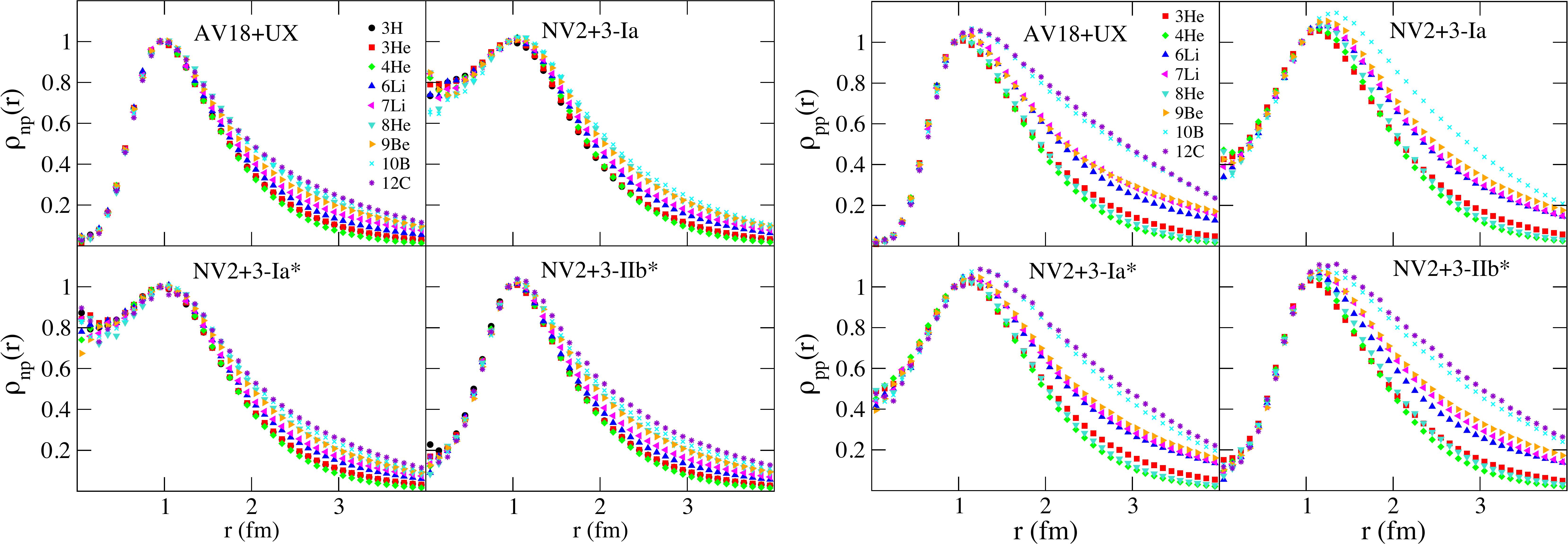}
 \captionsetup{justification=raggedright,singlelinecheck=false}
 \caption{Relative-distance $np$ (left panel) and $pp$ (right panel) pairs densities for several
nuclei. Distributions are shown for $^3$H, $^{3,4,8}$He, $^{6,7}$Li, $^9$Be, $^{10}$B, and $^{12}$C using the phenomenological AV18+UX and the local chiral NV2+3-Ia,  NV2+3-Ia*, and NV2+3-IIb* interactions. For each potential, all calculations are scaled to have the same value at $\sim 1$ fm. }  \label{2bdensities}
\end{figure*}

We can see that within a fixed interaction model, the two-nucleon densities at $r \lesssim 1.5$ fm for various nuclei exhibit a similar behavior, generated
by the cooperation of the short-range repulsion and the intermediate-range tensor attraction of the $N\!N$ interaction, with the tensor force governing the large overshoot at $r \sim 1.0$ fm between $np$ pairs.

As shown in Fig.~\ref{2bdensities}, where all calculations are scaled to have the same value at $\sim 1$ fm, the two-nucleon densities at short separations appears to be the same for all values of $A$, which leads to the nontrivial conclusion that at short ranges the two-nucleon motion is not affected by the presence of the other particles. 
This is what has been called universality of SRCs~\cite{Feldmeier:2011qy}. Moreover, at large separation the asymptotic behavior of the two-nucleon densities for different nuclei differs due to the different surface effects. 

While the short-distance behaviour is the same for all nuclei, it differs for each interaction. Indeed, the probability of finding two nucleons at short distances is finite for the "soft" NV2+3-Ia and NV2+3-Ia* chiral models, but approaches zero as we progress to the "hard" local chiral interaction NV2+3-IIb* and the "hardest" phenomenological AV18+UX. 



Nucleon pair distributions in different combinations of $ST$ for different nuclei can also be found online. In Table~\ref{tab:ST}, we report the number of pairs $N_{ST}$ for $^{3,4,6,8}$He, $^{6,7,8,9}$Li, $^{8,9,10}$Be, $^{10,11}$B, and $^{12}$C using the NV2+3 potentials. We show both independent pair (IP) numbers for the highest spatial symmetry states and for the fully correlated (cor) wave functions.  Correlated pair counts for the AV18+UX interaction fall within these ranges for all but two cases.

\begin{table}[t!]
\captionsetup{justification=raggedright,singlelinecheck=false}
\caption{Total number of spin-isospin $ST$ pairs in different nuclei for the NV2+3 potentials, showing both independent pair (IP) for the highest spatial symmetry states and the fully correlated (cor) wave functions.  Correlated pair counts for AV18+UX are consistent within error bars for all but a few cases.}
\setlength{\tabcolsep}{0.01\tabcolsep}
\begin{ruledtabular}
\begin{tabular}{ l c d d d d }
Nucleus & $\Psi$ & \multicolumn{1}{c}{$N_{01}$}
                 & \multicolumn{1}{c}{$N_{11}$}
                 & \multicolumn{1}{c}{$N_{10}$}
                 & \multicolumn{1}{c}{$N_{00}$} \\
\hline
$^3$He($\frac{1}{2}^+$)  & IP  & 1.5     & 0.0     & 1.5   & 0.0   \\
                         & cor & 1.37    & 0.13    & 1.49  & 0.01  \\
$^4$He($0^+$)            & IP  & 3.0     & 0.0     & 3.0   & 0.0   \\
                         & cor & 2.57(1) & 0.43(1) & 2.99  & 0.01  \\
$^6$He($0^+$)            & IP  & 5.5     & 4.5     & 4.5   & 0.5   \\
                         & cor & 4.95(1) & 5.05(1) & 4.49  & 0.51  \\
$^6$Li($1^+$)            & IP  & 4.5     & 4.5     & 5.5   & 0.5   \\
                         & cor & 4.07(2) & 4.93(2) & 5.47  & 0.53  \\
$^7$Li($\frac{3}{2}^-$)  & IP  & 6.75    & 6.75    & 6.75  & 0.75  \\
                         & cor & 6.13(2) & 7.37(2) & 6.73  & 0.77  \\
$^8$He($0^+$)            & IP  & 9.0     & 12.0    & 6.0   & 1.0   \\
                         & cor & 8.16(2) & 12.84(2)& 6.00  & 1.00  \\
$^8$Li($2^+$)            & IP  & 8.0     & 11.0    & 8.0   & 1.0   \\
                         & cor & 7.42(3) & 11.58(3)& 7.94  & 1.06  \\
$^8$Be($0^+$)            & IP  & 9.0     & 9.0     & 9.0   & 1.0   \\
                         & cor & 8.09(3) & 9.91(3) & 8.97  & 1.03  \\
$^9$Li($\frac{3}{2}^-$)  & IP  & 10.5    & 15.0    & 9.0   & 1.5   \\
                         & cor & 9.46(12)& 16.04(11)& 8.99(1)& 1.51(1)  \\
$^9$Be($\frac{3}{2}^-$)  & IP  & 10.5    & 13.5    & 10.5  & 1.5   \\
                         & cor & 9.57(4) & 14.43(4)& 10.46 & 1.54  \\ 
$^{10}$Be($0^+$)         & IP  & 13.0    & 18.0    & 12.0  & 2.0   \\
                         & cor & 11.73(4)& 19.27(4)& 11.98 & 2.02  \\
$^{10}$B($3^+$)          & IP  & 12.0    & 18.0    & 13.0  & 2.0   \\
                         & cor & 11.01(7)& 18.99(7)& 12.94(1) & 2.06(1)  \\
$^{11}$B($\frac{3}{2}^-$)& IP  & 15.0    & 22.5    & 15.0  & 2.5   \\
                         & cor & 13.84(2)& 23.66(3)& 14.91(3) & 2.59(3)  \\
$^{12}$C($0^+$)          & IP  & 18.0    & 27.0    & 18.0  & 3.0  \\
                         & cor & 16.54(6)& 28.46(6)& 17.91(1) & 3.09(1)
\label{tab:ST}
\end{tabular}
\end{ruledtabular}
\end{table}

A common feature in the $ST$ pair counts is that there is a moderate 10-15\% depletion of the $ST=01$ pairs going from IP to correlated wave functions, with a corresponding increase in the number of $ST=11$ pairs.  This is attributable to the many-body tensor correlations, which can flip spins (in exchange for orbital angular momentum) but not change isospin.  Because the $ST=01$ interactions are more attractive than $ST=11$, this depletion mechanism is a source of saturation of the nuclear binding.  The $ST=10$ pairs also show a depletion going from IP to correlated wave functions, with an increase of $ST=00$ pairs, but the effect is much smaller, probably because $ST=00$ interaction is generally much more repulsive than $ST=11$.

The probability of finding two nucleons with relative separation $r$ and center-of-mass distance $R$ is described by the calculation of the full probability density $\rho_{N\!N}(r,R)$. These densities are computationally demanding and are not available for all nuclei and interactions, but they can be generated upon request.

\begin{figure*}[b]
 \includegraphics[width=\linewidth]{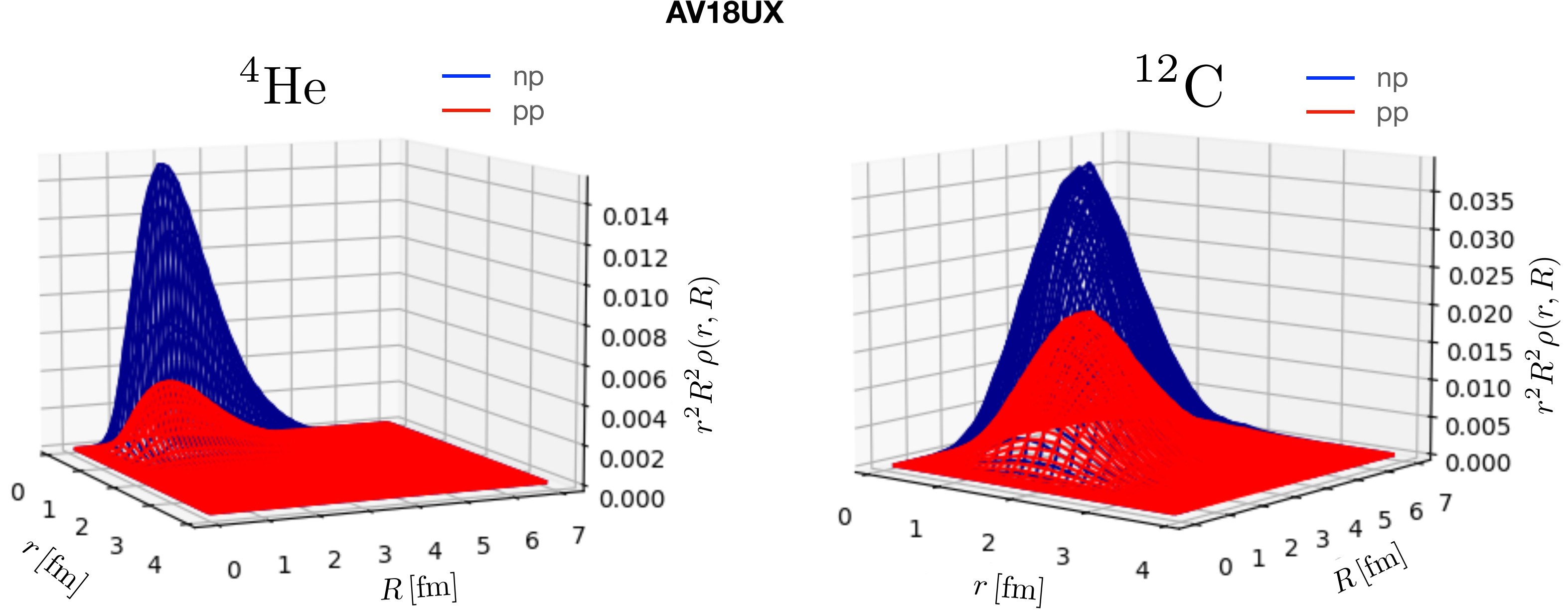}
 \captionsetup{justification=raggedright,singlelinecheck=false}
 \caption{The surface plots show the $np$ and $pp$ densities as function of the relative distance $r$ and center-of-mass $R$ for $^4$He (left panel) and $^{12}$C (right panel) using the phenomenological AV18+UX. The curves are normalized to obtain the corresponding $np$ and $pp$ pairs. }  \label{4He_12C_rhorR.av18ux.np.pp}
\end{figure*}

In Fig.~\ref{4He_12C_rhorR.av18ux.np.pp}, we present the the $np$ and $pp$ densities, multiplied by $r^2 R^2$, as a function of $r$ and $R$ for $^4$He and $^{12}$C using the phenomenological AV18+UX interaction. The curves are normalized to obtain the corresponding $np$ and $pp$ pairs, 4 $np$ and 1 $pp$ pairs in $^4$He,  36 $np$ and 15 $pp$ pairs in $^{12}$C.

\begin{figure*}[h]
 \includegraphics[width=\linewidth]{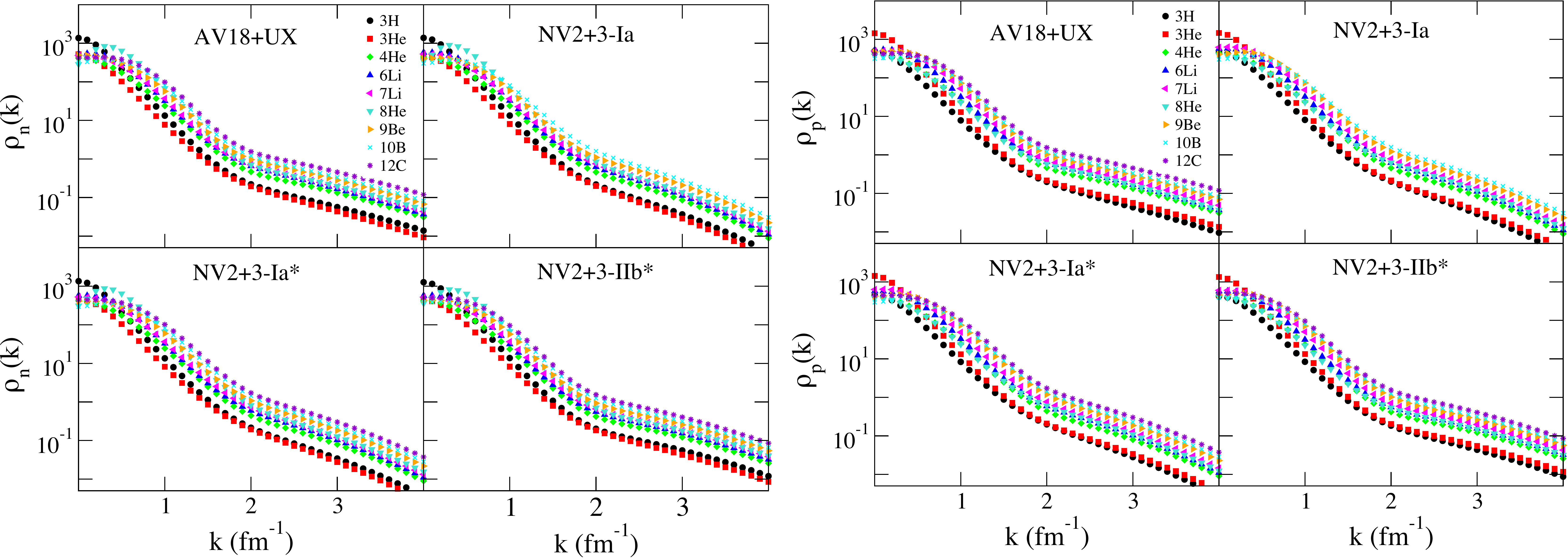}
 \captionsetup{justification=raggedright,singlelinecheck=false}
 \caption{Total one-body neutron (left panel) and proton (right panel) momentum distributions for $^3$H, $^3$He, $^4$He, $^6$Li, $^7$Li, $^8$He, $^9$Be, $^{10}$B, and $^{12}$C using the AV18+UIX phenomenological potentials, and the NV2+3-Ia,  NV2+3-Ia*, and NV2+3-IIb* local chiral interactions.}\label{1bmomentum}
\end{figure*}

\section{Momentum distributions}
\label{sec:mom}

\subsection {One-body momentum distributions}
\label{sec:single}
The probability of finding a nucleon with momentum $k$ and spin-isospin projection $\sigma$,$\tau$ in a given nuclear state is obtained by the Fourier transform of the one-nucleon nondiagonal density matrix

\begin{eqnarray}
\label{eq:rhok}
\rho_{\sigma\tau}({\bf k})\!\!&=&\!\!
\int d{\bf r}^\prime_1\, d{\bf r}_1\, d{\bf r}_2 \cdots d{\bf r}_A\,
\psi^\dagger_{JM_J}({\bf r}_1^\prime,{\bf r}_2, \dots,{\bf r}_A)\,  \nonumber \\
& \times &\, e^{-i{\bf k}\cdot ({\bf r}_1-{\bf r}^\prime_1)}
\, P_{\sigma\tau}(1) \, 
\psi_{JM_J} ({\bf r}_1,{\bf r}_2, \dots,{\bf r}_A) \, .
\end{eqnarray}

where $P_{\sigma\tau}(i)$ is the spin-isospin projection operator for nucleon $i$, and  $\psi_{JM_J}$ is the nuclear wave function with total spin $J$ and 
spin projection $M_J$.  The normalization is
\begin{equation}
\label{eq:nst}
  N_{\sigma \tau} = 
    \int \frac{d{\bf k}}{(2\pi)^3}\,\,\rho_{\sigma\tau}({\bf k}) \ ,
\end{equation}
where $N_{\sigma \tau}$ is the number of spin-up or spin-down protons
or neutrons.

Monte Carlo (MC) integration is used to construct the Fourier transform in Eq.~(\ref{eq:rhok}).  A conventional Metropolis walk, guided by 
$|\psi_{JM_J} ({\bf r}_1,\dots,{\bf r}_i,\dots,{\bf r}_A)|^2$, is used
to sample configurations~\cite{Pudliner:1997ck}.  
We average across all particles $i$ in each configuration, and for each
particle, the Fourier transform is computed using a grid of Gauss-Legendre points $\bf x_i$.
Instead of just moving the position ${\bf r}_i^\prime$ in the left-hand
wave function away from a fixed position ${\bf r}_i$ in the right-hand 
wave function, both positions are moved 
symmetrically away from ${\bf r}_i$, so Eq.~(\ref{eq:rhok}) becomes
\begin{eqnarray}
\label{eq:actual}
  \rho_{\sigma\tau}({\bf k}) &=& \frac{1}{A}\sum_i
  \int d{\bf r}_1 \cdots d{\bf r}_i \cdots d{\bf r}_A \int d{\bf \Omega}_x
  \int_0^{x_{\rm max}} x^2 dx \nonumber \\
  & & \psi^\dagger_{JM_J} 
  ({\bf r}_1,\dots,{\bf r}_i+{\bf x}/2,\dots,{\bf r}_A) \,
  e^{-i{\bf k}\cdot {\bf x} }  \\
  &\times& \,  P_{\sigma\tau}(i) \,
  \psi_{JM_J} 
  ({\bf r}_1,\dots,{\bf r}_i-{\bf x}/2,\dots,{\bf r}_A) \nonumber \, .
\end{eqnarray}
Here the polar angle $d{\bf \Omega}_x$ is also sampled by MC integration, with 
a randomly chosen direction for each particle in each MC configuration.
This approach is analogous to that used in studies of the nucleon-pair momentum distribution, see Refs.~\cite{Schiavilla:2006xx,Wiringa:2013ala}, and has the benefit of significantly decreasing statistical errors caused by the rapidly oscillating nature of the integrand for large values of $k$.  
To reach momenta $k \sim 10$ fm$^{-1}$ in $^4$He with good statistics 
requires integrating to $x_{\rm max}$=20 fm using 200 Gauss-Legendre points.\\

The results for a variety of nuclei in the range $A=2-12$ are available on the web page at \url{www.phy.anl.gov/theory/research/momenta/}. They are generated as distributions for neutron spin-down $\rho_{n\downarrow}(k)$, neutron spin-up $\rho_{n\uparrow}(k)$, proton spin-down $\rho_{p\downarrow}(k)$, and proton spin-up $\rho_{p\uparrow}(k)$, for the $M_J=J$ state. Where proton and neutron momentum distributions are the same, as in $T=0$ nuclei, only one set is given, and similarly, if spin-up and spin-down projections are the same, as in $0^{+}$ states, we give totals only.

In Fig.~\ref{1bmomentum} we show the total one-body neutron (left panel), $\rho_{n\downarrow}(k)+\rho_{n\uparrow}(k)$, and proton (right panel), $\rho_{p\downarrow}(k)+\rho_{p\uparrow}(k)$, momentum distributions for $^3$H, $^{3,4}$He, $^{6,7}$Li, $^8$He, $^9$Be, $^{10}$B, and $^{12}$C using the phenomenological AV18+UX, and the local chiral NV2+3-Ia,  NV2+3-Ia*, and NV2+3-IIb* interactions. Additional results for $^2$H, $^6$He, $^{8,9}$Li, $^{8,10}$Be, and $^{11}$B are shown in the online tables.

All models show the progressive high-momentum behavior in $k$ as the number of nucleons increases. Adding nucleons to the p-shell widens the distribution at low momenta and creates a peak at a finite $k$. All of these nuclei have a dramatic shift in slope at $k = 2$ fm$^{-1}$ to a broad shoulder, which is attributed to the large tensor correlation caused by the pion-exchange component of the nuclear force. As expected, the differences between the models are most noticeable in the high-momentum tails, which decay more rapidly with increasing $k$ for the "soft" NV2+3-Ia and NV2+3-Ia* interactions than the "hard" NV2+3-IIb* and AV18+UX potentials. The difference in momentum distributions observed for the NV2+3-Ia and NV2+3-Ia* models, which have the same two-body interaction but different the three-nucleon force parametrization, is small. For NV2+3-Ia, there is a minor overall shift in $\rho_n(k)$ and $\rho_p(k)$ toward bigger $k$.

\subsection {Two-body momentum distributions}
\label{sec:double}

The probability of finding two nucleons in a nucleus with relative momentum
${\bf q}=({\bf k}_1-{\bf k}_2)/2$ and total center-of-mass momentum
${\bf Q}={\bf k}_1+{\bf k}_2$ in a given spin-isospin state is given by:
\begin{eqnarray}
\label{eq:rhoqQ}
\rho_{ST}({\bf q},{\bf Q}) 
&=&\int d{\bf r}^\prime_1 d{\bf r}_1 d{\bf r}^\prime_2d{\bf r}_2 d{\bf r}_3 \cdots d{\bf r}_A \nonumber \\
&& \psi^\dagger_{JM_J}({\bf r}_1^\prime,{\bf r}^\prime_2,{\bf r}_3,\dots,{\bf r}_A)  \\
&& e^{-i{\bf q}\cdot({\bf r}-{\bf r}^\prime)} 
e^{-i{\bf Q}\cdot({\bf R}-{\bf R}^\prime)} \nonumber \\
&&  P_{ST}(12) \psi_{JM_J}({\bf r}_1,{\bf r}_2,{\bf r}_3,\dots,{\bf r}_A) \, ,
\nonumber
\end{eqnarray}
where ${\bf r} = {\bf r}_1 - {\bf r}_2$,
${\bf R} = ({\bf r}_1 + {\bf r}_2)/2$,
and $P_{ST}(12)$ is a projector onto pair spin $S=0$ or 1, and isospin
$T=0$ or 1.
The total normalization is:
\begin{equation}
N_{ST} = \int \frac{d{\bf q}}{(2\pi)^3} \frac{d{\bf Q}}{(2\pi)^3}
         \rho_{ST}({\bf q},{\bf Q}) \, ,
\end{equation}
where $N_{ST}$ is the total number of nucleon pairs with given spin-isospin.
Alternate projectors can also be used, e.g., for $N\!N$ pairs $pp$, $np$, 
and $nn$ (and each of these with spin S) with corresponding normalizations.

The nucleon-pair momentum distributions can be examined in a number of different ways.
One way is to integrate over all values of ${\bf Q}$ and reduce the total pair
density to a function $\rho_{ N\!N}(q)$ of the relative momentum $q$ only.
In this case, Eq.(\ref{eq:rhoqQ}) reduces to a form similar to
Eq.(\ref{eq:actual}), with a sum over all configurations in the Monte
Carlo walk controlled by $|\Psi_{JM_J}|^2$, and a Gauss-Legendre integration
over the relative separation ${\bf x} = {\bf r}-{\bf r}^\prime$.
Again, the polar angle ${\bf \Omega}_x$ is sampled by randomly choosing the
direction of ${\bf x}$ in space, and an average over all pairs in every
MC configuration is made.

Many results for $\rho_{ST}(q)$ and $\rho_{N\!N}(q)$ obtained for various light nuclei in the range $A=3-12$ are recorded in the online tables at \url{www.phy.anl.gov/theory/research/momenta2/}. These are from VMC calculations using different Norfolk NV2+3 potentials, including -Ia, -Ia*, -Ib*, -IIa*, and -IIb*, as well as results obtained with the AV18+UX.
The nuclei covered include $^3$H, $^{3,4,6,8}$He, $^{6,7,8,9}$Li, $^{8,9}$Be, $^{10}$B, and $^{12}$C.

\begin{figure*}[b]
 \includegraphics[width=\linewidth]{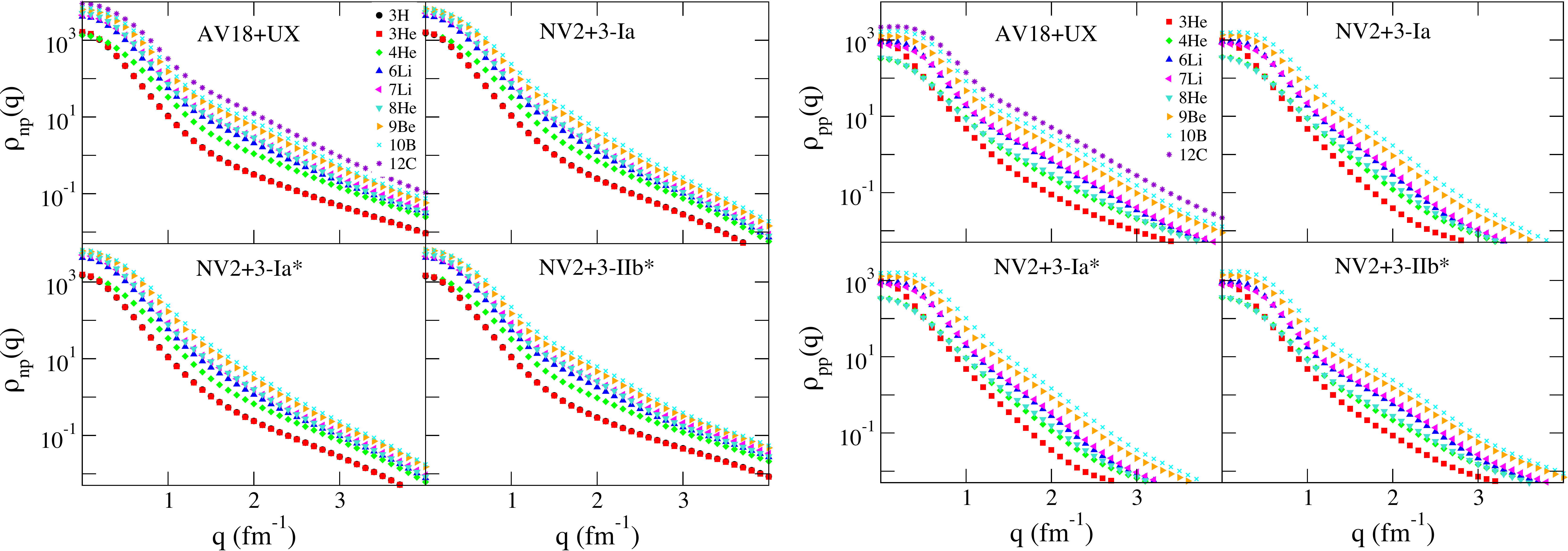}
 \captionsetup{justification=raggedright,singlelinecheck=false}
 \caption{Total $np$ (left panel) and $pp$ (right panel) momentum distributions as function of the relative momentum $q$ for $^3$H, $^3$He, $^4$He, $^6$Li, $^7$Li, $^8$He, $^9$Be, $^{10}$B, and $^{12}$C using the AV18+UIX phenomenological potentials, and the NV2+3-Ia,  NV2+3-Ia*, and NV2+3-IIb* local chiral interactions.}  \label{2bmomentum}
\end{figure*}

In Fig.~\ref{2bmomentum}, we display the $np$ and $pp$ momentum distributions for selected nuclei using the AV18+UX phenomenological potentials, and the NV2+3-Ia,  NV2+3-Ia*, and NV2+3-IIb* local chiral interactions. They have been calculated for relative momentum $q$ from 0 to 10\, fm$^{-1}$ and integrated over all values of ${\bf Q}$. All the four Hamiltonians show the high-momentum tail in $q$, but it decays more rapidly for the soft NV2+3-Ia and NV2+3-Ia*.

\begin{figure*}[t]
 \includegraphics[width=\linewidth]{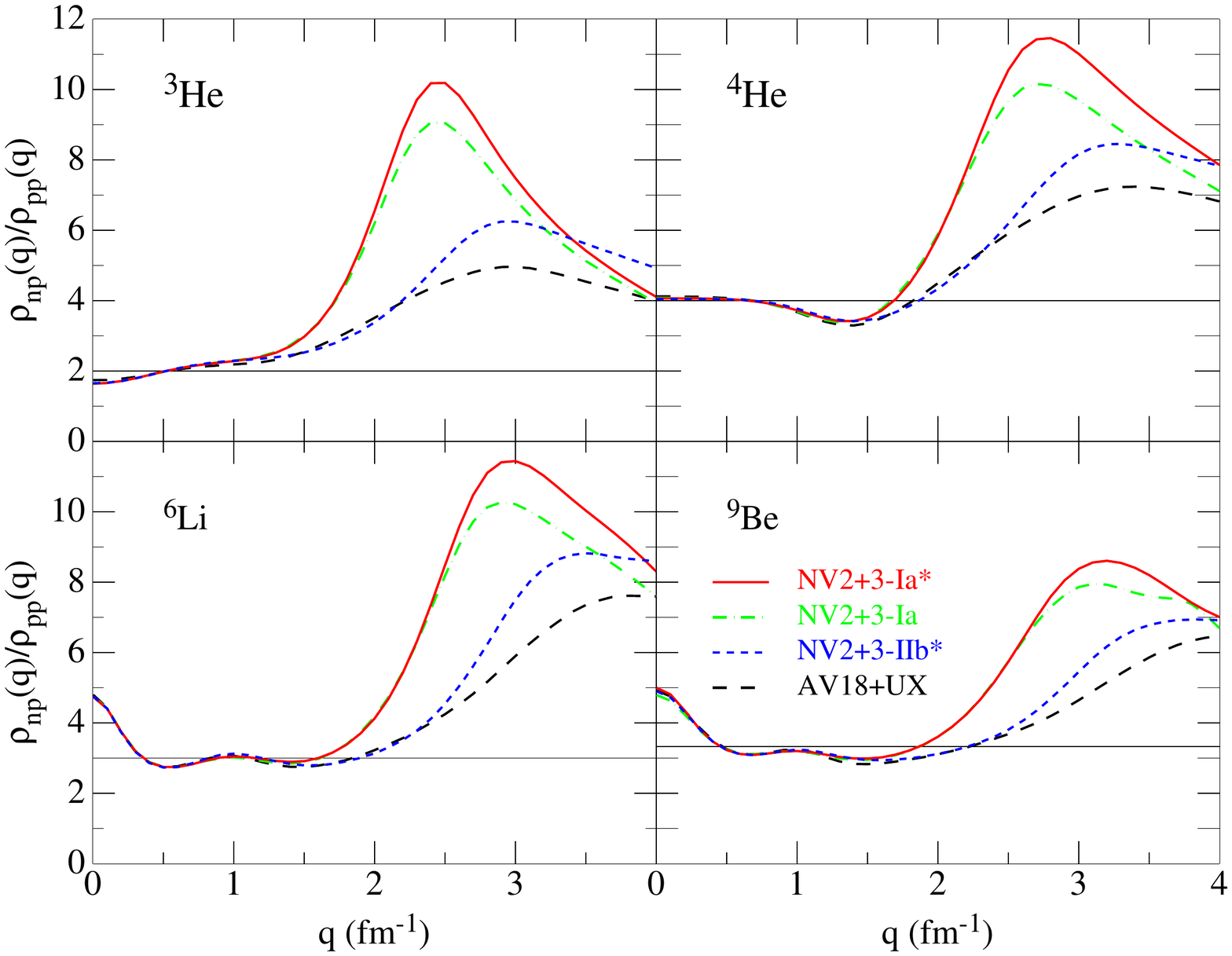}
 \caption{Ratio of $np$ to $pp$ pairs as function of relative momentum $q$
 for various nuclei; the constant line in each panel is the total pair ratio.}  \label{ratio}
\end{figure*}

In Fig.~\ref{ratio}, we show the ratio of the $np$ and $pp$ momentum distributions for several nuclei relative to the total $np/pp$ ratio.  For $q\leq 2$ fm$^{-1}$ the ratios are virtually identical for the different interactions and close to the total number of $np/pp$ pairs.  Beyond that point, the $np/pp$ ratio gets larger, with the soft interactions showing a larger peak at smaller $q$, while the hard interactions interactions show a lower but broader peak at larger $q$.  This behavior is probably due to the strong tensor correlations in the $np$ channel.
Note these ratios are integrated over all $Q$, while the much larger $np/pp$ ratios, mentioned above from Refs.~\cite{Hen:2014nza,LabHallA:2014wqo,CLAS:2018xvc} are for small $Q$.

\begin{figure*}[h]
 \includegraphics[width=\linewidth]{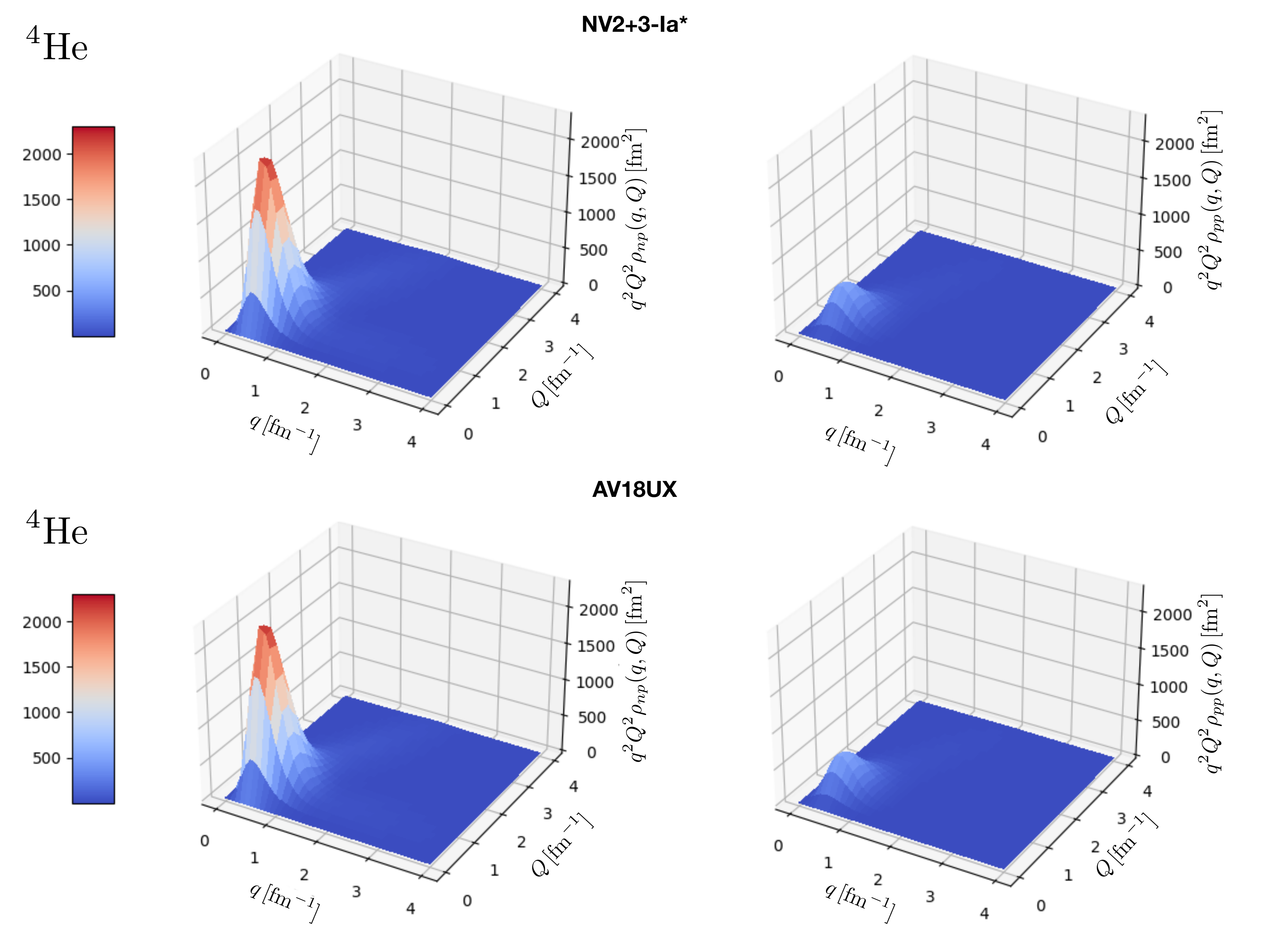}
 \captionsetup{justification=raggedright,singlelinecheck=false}
 \caption{The surface plots show the the $np$ (left panels) and $pp$ (right panels) momentum distributions as functions of the relative momentum $q$ and center-of-mass momentum $Q$ for the alpha particle obtained with the NV2+3-Ia* and AV18+UX interactions.}    \label{He.rho(q,Q)_tot}
\end{figure*}

\begin{figure*}[h]
 \includegraphics[width=\linewidth]{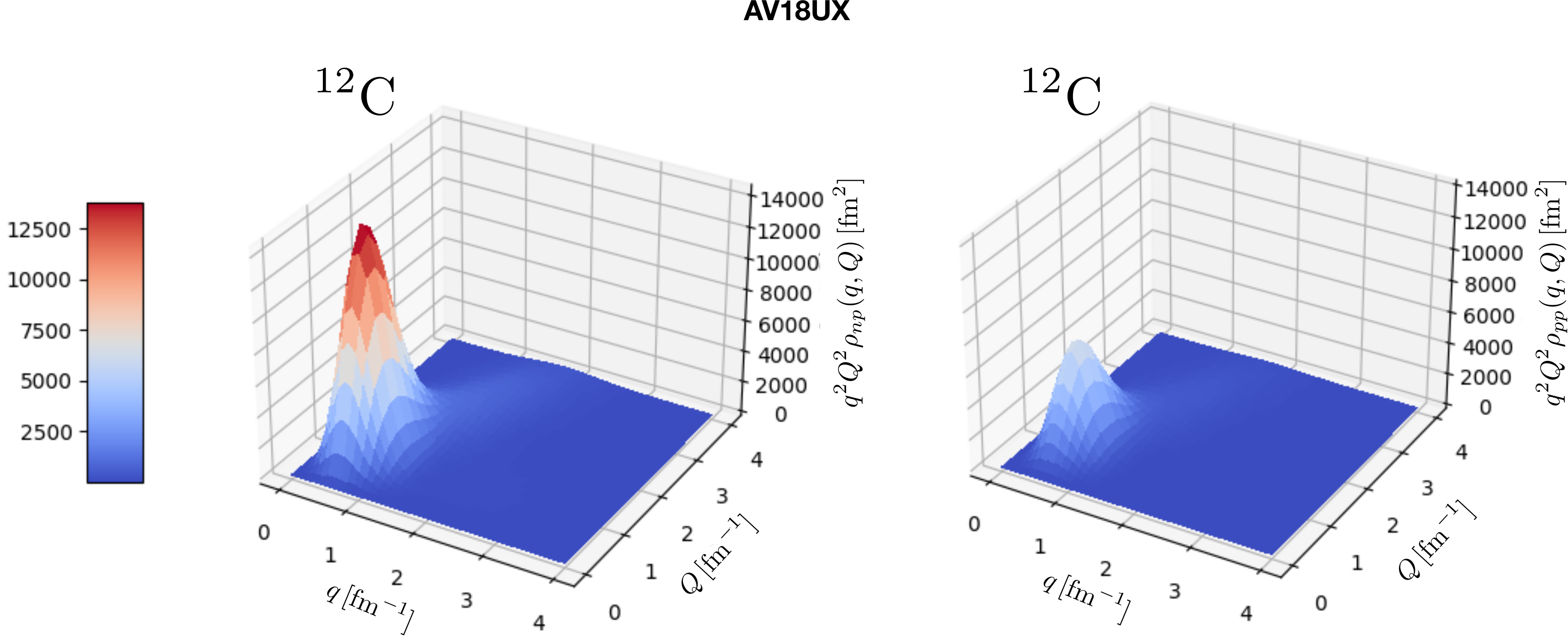}
 \captionsetup{justification=raggedright,singlelinecheck=false}
 \caption{Same as Fig.~\ref{He.rho(q,Q)_tot} but for $^{12}$C using the AV18+UX. }  \label{C.rho(q,Q)_tot}
\end{figure*}

We can also integrate Eq.(\ref{eq:rhoqQ}) over all ${\bf q}$, leaving a function $\rho_{N\!N}(Q)$ of the total pair momentum $Q$ only.  In general, the $\rho_{N\!N}(Q)$ for a given nucleus has a smaller falloff at large momenta than the $\rho_{N\!N}(q)$ and the ratios of different $N\!N$ components vary less over the range of $Q$. These distributions are generally not as interesting, but they can be generated on request.

Calculations of the full $\rho_{ N\!N}({\bf q},{\bf Q})$ are more challenging as they require a double Gauss-Legendre integral over two randomly chosen directions for each pair in each MC sample: ${\bf x}$ for ${\bf r}$ and ${\bf X}$ for ${\bf R}$. In Fig.~\ref{He.rho(q,Q)_tot}, we display the surface plots of $\rho_{np}({\bf q},{\bf Q})$ and $\rho_{pp}({\bf q},{\bf Q})$ as functions of the relative momentum $q$ and center-of-mass momentum $Q$ for $^4$He using the NV2+3-Ia* and AV18+UX interactions.  In Fig.~\ref{C.rho(q,Q)_tot} we show similar plots for $^{12}$C but only for the AV18+UX interaction.  In the online tables we present $\rho_{ N\!N}({\bf q},{\bf Q})$ results for $^{3,4}$He, $^6$Li, $^{12}$Be, and $^{12}$C for the AV18+UX and one or both of the NV2+3-Ia* and NV2+3-IIb* interactions.

In addition, we can differentiate between short-range (SR) and long-range (LR) pair contributions by simply sorting our MC samples into two sets, where $r<a$ for SR pairs and $r>a$ for LR pairs. For example, in $^4$He, a boundary of $a=2$ fm divides the six $N\!N$ pairs into approximately two equal groups. Fig.~\ref{He.rho(q,Q)_2.0} shows the $np$ (left panels) and $pp$ (right panels) SR and LR pair distributions for the alpha particle using the NV2+3-Ia* and AV18+UX interactions. Taking the integral of these distributions we find that in the case of the NV2+3-Ia* the number of $np$ SR and LR pairs are 2.04 and 1.96, respectively, while the number of $pp$ SR and LR pairs are 0.53 and 0.47. For the case of the AV18+UX the of $np$ SR and LR pairs are 2.06 and 1.94 while the number of $pp$ SR and LR pairs are 0.48 and 0.52.

\begin{figure*}[h]
 \includegraphics[width=\linewidth]{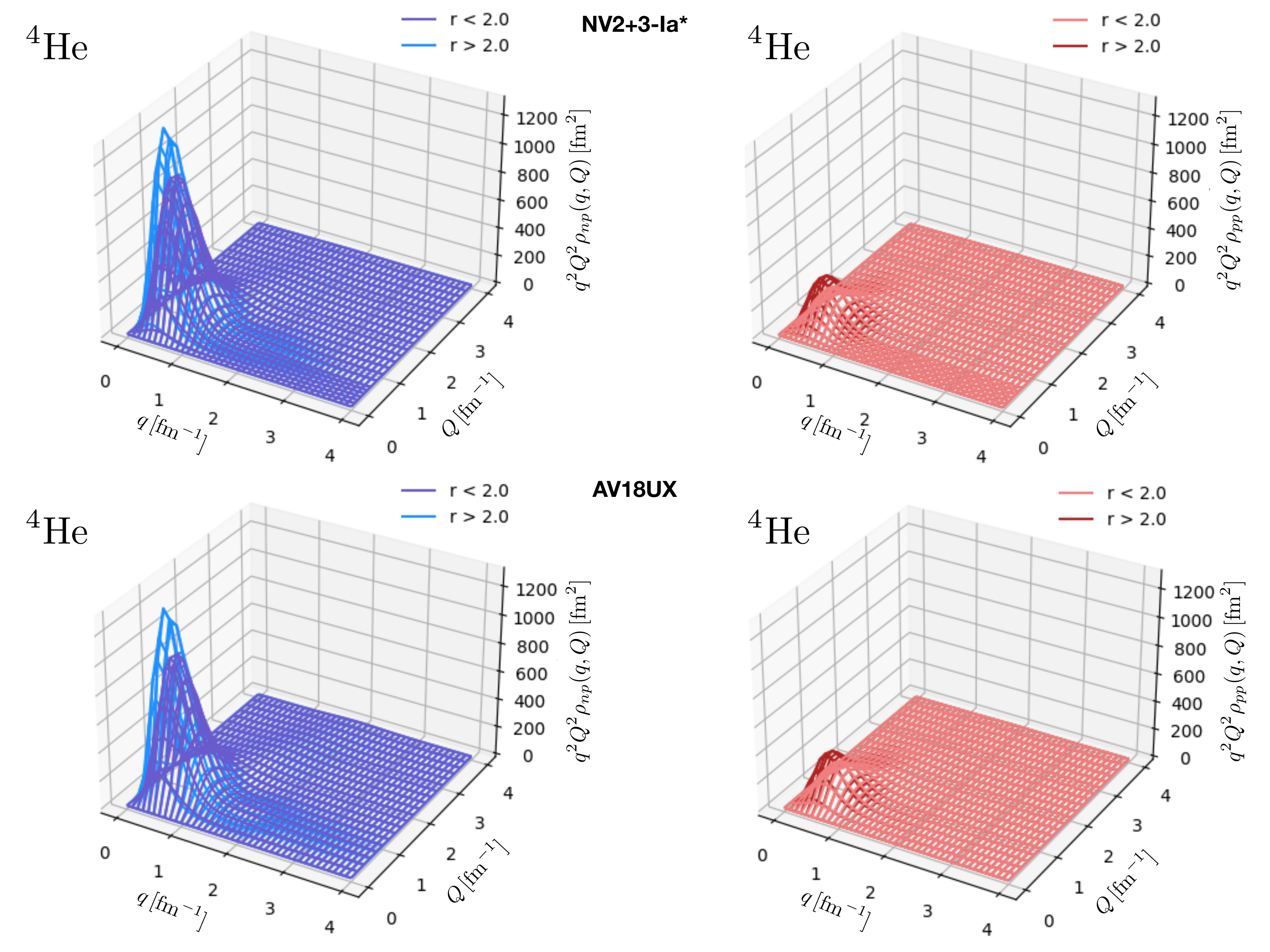}
 \caption{ $np$ (left panel) and $pp$ (right panel) momentum distributions in $^4$He from pairs with separation $r \leq 2.0$ fm of each other and from pairs with $r \geq 2.0$ fm. }  \label{He.rho(q,Q)_2.0}
\end{figure*}

Similarly, The $np$ and $pp$ SR and LR pair distributions for $^{12}$C employing the AV18+UX interactions are shown in Fig.~\ref{12C.rho(q,Q)_av18ux.2.5} for $a=2.5$ fm. By integrating these distributions, we find that the number of $np$ SR and LR pairings is 12.9 and 23.1 respectively, while the number of $pp$ SR and LR pairs is 4.3 and 10.7, respectively.
In the online figures and tables we also provide the breakdown for $^{3,4}$He and $^6$Li for the AV18+UX, NV2+3-Ia*, and NV2+3-IIb* with the break point $a=2.5$ fm.
These figures clearly show that the LR pairs dominate at low pair momenta but fall off rapidly beyond $q \approx 1.5$ fm$^{-1}$, while the SR pairs provide the high-momentum tail.  For the pair center of mass momentum, the total number of pairs declines significantly beyond $Q \approx 2$ fm$^{-1}$ but there continues to be a high-momentum tail in $q$.

\begin{figure*}[h]
 \includegraphics[width=\linewidth]{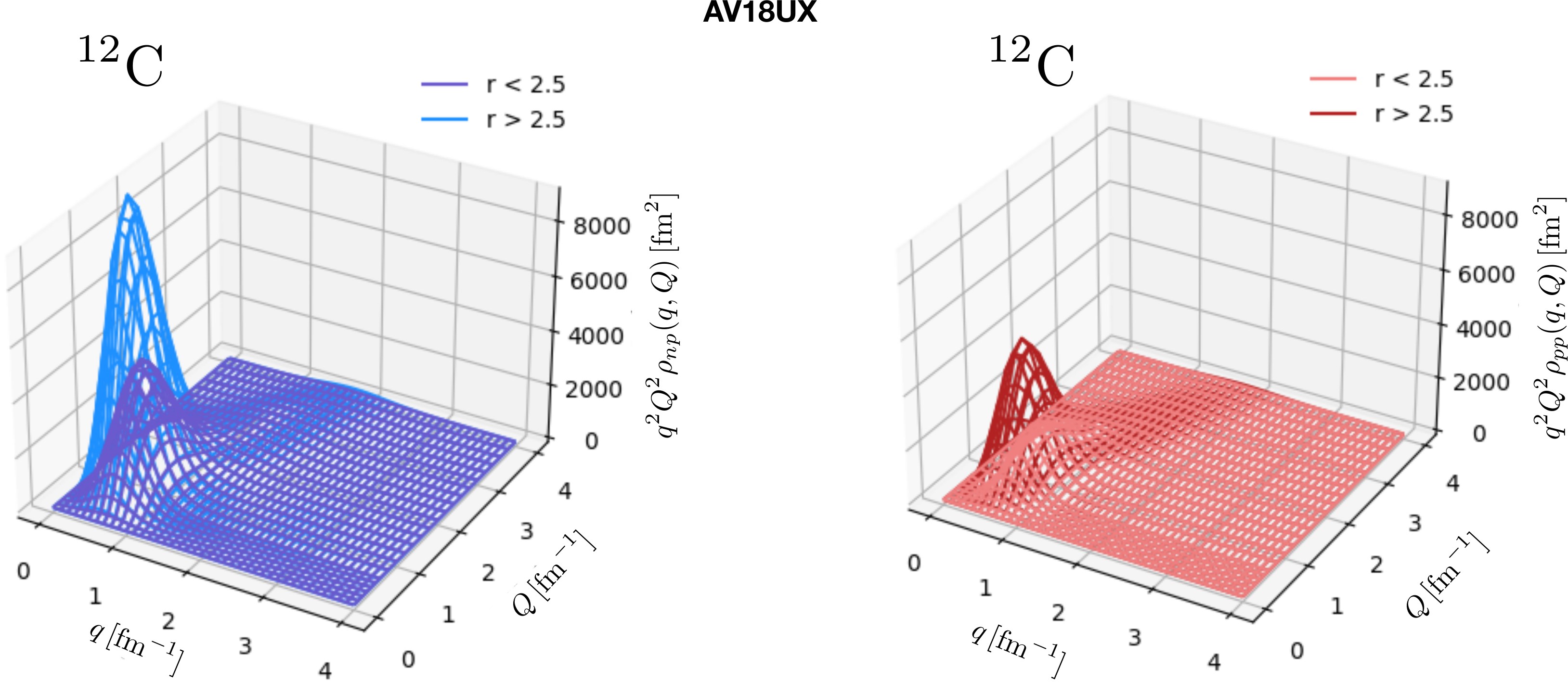}
 \caption{ Same as Fig.~\ref{He.rho(q,Q)_2.0} but for $^{12}$C using the AV18+UX with separation boundary $a = 2.5$ fm. }  \label{12C.rho(q,Q)_av18ux.2.5}
\end{figure*}

\section {Conclusions}
\label{sec:conclude}
We have performed VMC calculations of one- and two-body density distributions and one- and two-body momentum distributions for a wide variety of nuclei from $^2$H up to $^{12}$C, using the $\Delta$-full Norfolk interactions obtained from $\chi$EFT.
Results are compared to those obtained with the conventional AV18+UX interaction, some of which were previously reported in Ref.~\cite{Wiringa:2013ala}.
New features in the present work include i) calculations of the pair density $\rho_{N\!N}(r,R)$ as a function of both the pair separation $r$ and pair center-of-mass $R$; and ii) calculations of the two-body momentum distribution $\rho_{N\!N}(q,Q,a)$ coming from short- and long-range pairs differentiated by a pair separation boundary $a$.

Comparing results among the different NV2+3 and AV18+UX interactions, we find the one-body densities $\rho_N(r)$ for a given nucleus are very similar for all cases.  Also, the total number of spin-up/down protons and neutrons is remarkably constant.
In contrast, the two-body densities $\rho_{N\!N}(r)$ vary significantly at short distances, depending on whether the interaction is `soft' like NV2+3-Ia or `hard' like NV2+3-IIb*.  However, the total number of spin-isospin $ST$ pairs for a given nucleus shows little variation among the different interactions.

One-body momentum distributions $\rho(k)$ all share the same characteristics, with a maximum at $k=0$ fm$^{-1}$, a rapid fall off to $k\approx 2$ fm$^{-1}$, followed by a high-momentum tail that is more prominent for `hard' interactions, and less so for `soft' interactions.  For a given interaction, the low-$k$ behavior varies with the nucleus, but the high-$k$ tails are essentially parallel for all $A=2$-12 nuclei.  Two-body momentum distributions $\rho_{N\!N}(q)$ are similar, but tend to change slope near $q\approx 1.5$ fm$^{-1}$.  Again, the high-$q$ tail is larger for `hard' interactions.  We also note that the ratio $\rho_{np}(q)/\rho_{pp}(q)$ is relatively flat and proportional to the number of pairs of each type at lower $q$, but much larger at higher $q$, reflecting the importance of the stronger tensor correlations in $np$ versus $pp$ pairs.  Finally, our studies of $\rho_{N\!N}(q,Q,a)$, which separates the contributions of short- from long-range pairs, clearly indicates the high-$q$ tails are due to SRCs.

Concurrent to these studies, are QMC studies of nuclear electroweak response densities and response functions~\cite{Pastore:2019urn,Barrow:2020mfy,Andreoli:2021cxo} where the interaction of the external probes (both electron and neutrinos) is accounted for at one- and two-body level along with SRCs. In particular, within the Short-Time-Approximation~\cite{Pastore:2019urn} it is possible to analyze electroweak nuclear responses in terms of the kinematic variables, that is relative and center of mass momenta,  associated with a pair of correlated nucleons struck by the probe. Many-body effects in the coupling of electroweak probes with correlated nucleons are being vigorously investigated due to their relevance to both electron and neutrino scattering processes.  

While this paper provides examples of the densities and momentum distributions, the full set of results is accessible in graphical and tabular forms online at \url{www.phy.anl.gov/theory/research/QMCresults.html}.
We expect to continue expanding and updating these results in future.

\section*{Acknowledgments}
This work is supported by the U.S.~Department of Energy under contract DE-SC0021027 (S.~P.), DE-AC02-06CH11357 (R.~B.~W.), a 2021 Early Career Award number DE-SC0022002 (M.~P.), the FRIB Theory Alliance award DE-SC0013617 (S.~P. and M.~P.), and the NUCLEI SciDAC program (R.~B.~W.).

The many-body calculations were performed on the parallel computers of the Laboratory Computing Resource Center, Argonne National Laboratory, and the computers of 
the Argonne Leadership Computing Facility via the INCITE grant ``Ab-initio nuclear structure and nuclear reactions'', the 2019/2020 ALCC grant ``Low Energy Neutrino-Nucleus interactions'' for the project NNInteractions,
the 2020/2021 ALCC grant ``Chiral Nuclear Interactions from Nuclei to Nucleonic Matter'' for the project ChiralNuc, and by the 2021/2022 ALCC grant ``Quantum Monte Carlo Calculations of Nuclei up to $^{16}{\rm O}$ and Neutron Matter" for the project QMCNuc. 

%
%

\bibliographystyle{apsrev.bst}
\bibliography{biblio}

\begin{thebibliography}{50}
\expandafter\ifx\csname natexlab\endcsname\relax\def\natexlab#1{#1}\fi
\expandafter\ifx\csname bibnamefont\endcsname\relax
  \def\bibnamefont#1{#1}\fi
\expandafter\ifx\csname bibfnamefont\endcsname\relax
  \def\bibfnamefont#1{#1}\fi
\expandafter\ifx\csname citenamefont\endcsname\relax
  \def\citenamefont#1{#1}\fi
\expandafter\ifx\csname url\endcsname\relax
  \def\url#1{\texttt{#1}}\fi
\expandafter\ifx\csname urlprefix\endcsname\relax\def\urlprefix{URL }\fi
\providecommand{\bibinfo}[2]{#2}
\providecommand{\eprint}[2][]{\url{#2}}

\bibitem[{\citenamefont{Tang et~al.}(2003)}]{Tang:2002ww}
\bibinfo{author}{\bibfnamefont{A.}~\bibnamefont{Tang}} \bibnamefont{et~al.},
  \bibinfo{journal}{Phys. Rev. Lett.} \textbf{\bibinfo{volume}{90}},
  \bibinfo{pages}{042301} (\bibinfo{year}{2003}), \eprint{nucl-ex/0206003}.

\bibitem[{\citenamefont{Piasetzky et~al.}(2006)\citenamefont{Piasetzky,
  Sargsian, Frankfurt, Strikman, and Watson}}]{Piasetzky:2006ai}
\bibinfo{author}{\bibfnamefont{E.}~\bibnamefont{Piasetzky}},
  \bibinfo{author}{\bibfnamefont{M.}~\bibnamefont{Sargsian}},
  \bibinfo{author}{\bibfnamefont{L.}~\bibnamefont{Frankfurt}},
  \bibinfo{author}{\bibfnamefont{M.}~\bibnamefont{Strikman}}, \bibnamefont{and}
  \bibinfo{author}{\bibfnamefont{J.~W.} \bibnamefont{Watson}},
  \bibinfo{journal}{Phys. Rev. Lett.} \textbf{\bibinfo{volume}{97}},
  \bibinfo{pages}{162504} (\bibinfo{year}{2006}), \eprint{nucl-th/0604012}.

\bibitem[{\citenamefont{Shneor et~al.}(2007)}]{JeffersonLabHallA:2007lly}
\bibinfo{author}{\bibfnamefont{R.}~\bibnamefont{Shneor}} \bibnamefont{et~al.}
  (\bibinfo{collaboration}{Jefferson Lab Hall A}), \bibinfo{journal}{Phys. Rev.
  Lett.} \textbf{\bibinfo{volume}{99}}, \bibinfo{pages}{072501}
  (\bibinfo{year}{2007}), \eprint{nucl-ex/0703023}.

\bibitem[{\citenamefont{Subedi et~al.}(2008)}]{Subedi:2008zz}
\bibinfo{author}{\bibfnamefont{R.}~\bibnamefont{Subedi}} \bibnamefont{et~al.},
  \bibinfo{journal}{Science} \textbf{\bibinfo{volume}{320}},
  \bibinfo{pages}{1476} (\bibinfo{year}{2008}), \eprint{0908.1514}.

\bibitem[{\citenamefont{Fomin et~al.}(2012)}]{Fomin:2011ng}
\bibinfo{author}{\bibfnamefont{N.}~\bibnamefont{Fomin}} \bibnamefont{et~al.},
  \bibinfo{journal}{Phys. Rev. Lett.} \textbf{\bibinfo{volume}{108}},
  \bibinfo{pages}{092502} (\bibinfo{year}{2012}), \eprint{1107.3583}.

\bibitem[{\citenamefont{Korover et~al.}(2014)}]{LabHallA:2014wqo}
\bibinfo{author}{\bibfnamefont{I.}~\bibnamefont{Korover}} \bibnamefont{et~al.}
  (\bibinfo{collaboration}{Lab Hall A}), \bibinfo{journal}{Phys. Rev. Lett.}
  \textbf{\bibinfo{volume}{113}}, \bibinfo{pages}{022501}
  (\bibinfo{year}{2014}), \eprint{1401.6138}.

\bibitem[{\citenamefont{Hen et~al.}(2014)}]{Hen:2014nza}
\bibinfo{author}{\bibfnamefont{O.}~\bibnamefont{Hen}} \bibnamefont{et~al.},
  \bibinfo{journal}{Science} \textbf{\bibinfo{volume}{346}},
  \bibinfo{pages}{614} (\bibinfo{year}{2014}), \eprint{1412.0138}.

\bibitem[{\citenamefont{Duer et~al.}(2019)}]{CLAS:2018xvc}
\bibinfo{author}{\bibfnamefont{M.}~\bibnamefont{Duer}} \bibnamefont{et~al.}
  (\bibinfo{collaboration}{CLAS}), \bibinfo{journal}{Phys. Rev. Lett.}
  \textbf{\bibinfo{volume}{122}}, \bibinfo{pages}{172502}
  (\bibinfo{year}{2019}), \eprint{1810.05343}.

\bibitem[{\citenamefont{Duer et~al.}(2018)}]{CLAS:2018yvt}
\bibinfo{author}{\bibfnamefont{M.}~\bibnamefont{Duer}} \bibnamefont{et~al.}
  (\bibinfo{collaboration}{CLAS}), \bibinfo{journal}{Nature}
  \textbf{\bibinfo{volume}{560}}, \bibinfo{pages}{617} (\bibinfo{year}{2018}).

\bibitem[{\citenamefont{Schiavilla et~al.}(2007)\citenamefont{Schiavilla,
  Wiringa, Pieper, and Carlson}}]{Schiavilla:2006xx}
\bibinfo{author}{\bibfnamefont{R.}~\bibnamefont{Schiavilla}},
  \bibinfo{author}{\bibfnamefont{R.~B.} \bibnamefont{Wiringa}},
  \bibinfo{author}{\bibfnamefont{S.~C.} \bibnamefont{Pieper}},
  \bibnamefont{and} \bibinfo{author}{\bibfnamefont{J.}~\bibnamefont{Carlson}},
  \bibinfo{journal}{Phys. Rev. Lett.} \textbf{\bibinfo{volume}{98}},
  \bibinfo{pages}{132501} (\bibinfo{year}{2007}), \eprint{nucl-th/0611037}.

\bibitem[{\citenamefont{Alvioli et~al.}(2008)\citenamefont{Alvioli, Ciofi~degli
  Atti, and Morita}}]{Alvioli:2007zz}
\bibinfo{author}{\bibfnamefont{M.}~\bibnamefont{Alvioli}},
  \bibinfo{author}{\bibfnamefont{C.}~\bibnamefont{Ciofi~degli Atti}},
  \bibnamefont{and} \bibinfo{author}{\bibfnamefont{H.}~\bibnamefont{Morita}},
  \bibinfo{journal}{Phys. Rev. Lett.} \textbf{\bibinfo{volume}{100}},
  \bibinfo{pages}{162503} (\bibinfo{year}{2008}).

\bibitem[{\citenamefont{Feldmeier et~al.}(2011)\citenamefont{Feldmeier,
  Horiuchi, Neff, and Suzuki}}]{Feldmeier:2011qy}
\bibinfo{author}{\bibfnamefont{H.}~\bibnamefont{Feldmeier}},
  \bibinfo{author}{\bibfnamefont{W.}~\bibnamefont{Horiuchi}},
  \bibinfo{author}{\bibfnamefont{T.}~\bibnamefont{Neff}}, \bibnamefont{and}
  \bibinfo{author}{\bibfnamefont{Y.}~\bibnamefont{Suzuki}},
  \bibinfo{journal}{Phys. Rev. C} \textbf{\bibinfo{volume}{84}},
  \bibinfo{pages}{054003} (\bibinfo{year}{2011}), \eprint{1107.4956}.

\bibitem[{\citenamefont{Alvioli et~al.}(2013)\citenamefont{Alvioli, Ciofi~degli
  Atti, Kaptari, Mezzetti, and Morita}}]{Alvioli:2012qa}
\bibinfo{author}{\bibfnamefont{M.}~\bibnamefont{Alvioli}},
  \bibinfo{author}{\bibfnamefont{C.}~\bibnamefont{Ciofi~degli Atti}},
  \bibinfo{author}{\bibfnamefont{L.~P.} \bibnamefont{Kaptari}},
  \bibinfo{author}{\bibfnamefont{C.~B.} \bibnamefont{Mezzetti}},
  \bibnamefont{and} \bibinfo{author}{\bibfnamefont{H.}~\bibnamefont{Morita}},
  \bibinfo{journal}{Phys. Rev. C} \textbf{\bibinfo{volume}{87}},
  \bibinfo{pages}{034603} (\bibinfo{year}{2013}), \eprint{1211.0134}.

\bibitem[{\citenamefont{Rios et~al.}(2014)\citenamefont{Rios, Polls, and
  Dickhoff}}]{Rios:2013zqa}
\bibinfo{author}{\bibfnamefont{A.}~\bibnamefont{Rios}},
  \bibinfo{author}{\bibfnamefont{A.}~\bibnamefont{Polls}}, \bibnamefont{and}
  \bibinfo{author}{\bibfnamefont{W.~H.} \bibnamefont{Dickhoff}},
  \bibinfo{journal}{Phys. Rev. C} \textbf{\bibinfo{volume}{89}},
  \bibinfo{pages}{044303} (\bibinfo{year}{2014}), \eprint{1312.7307}.

\bibitem[{\citenamefont{Ciofi~degli Atti et~al.}(1991)\citenamefont{Ciofi~degli
  Atti, Simula, Frankfurt, and Strikman}}]{CiofidegliAtti:1991mm}
\bibinfo{author}{\bibfnamefont{C.}~\bibnamefont{Ciofi~degli Atti}},
  \bibinfo{author}{\bibfnamefont{S.}~\bibnamefont{Simula}},
  \bibinfo{author}{\bibfnamefont{L.~L.} \bibnamefont{Frankfurt}},
  \bibnamefont{and} \bibinfo{author}{\bibfnamefont{M.~I.}
  \bibnamefont{Strikman}}, \bibinfo{journal}{Phys. Rev. C}
  \textbf{\bibinfo{volume}{44}}, \bibinfo{pages}{R7} (\bibinfo{year}{1991}).

\bibitem[{\citenamefont{Wiringa et~al.}(2014)\citenamefont{Wiringa, Schiavilla,
  Pieper, and Carlson}}]{Wiringa:2013ala}
\bibinfo{author}{\bibfnamefont{R.~B.} \bibnamefont{Wiringa}},
  \bibinfo{author}{\bibfnamefont{R.}~\bibnamefont{Schiavilla}},
  \bibinfo{author}{\bibfnamefont{S.~C.} \bibnamefont{Pieper}},
  \bibnamefont{and} \bibinfo{author}{\bibfnamefont{J.}~\bibnamefont{Carlson}},
  \bibinfo{journal}{Phys. Rev. C} \textbf{\bibinfo{volume}{89}},
  \bibinfo{pages}{024305} (\bibinfo{year}{2014}), \eprint{1309.3794}.

\bibitem[{\citenamefont{Ciofi~degli Atti et~al.}(2017)\citenamefont{Ciofi~degli
  Atti, Mezzetti, and Morita}}]{CiofidegliAtti:2017tnm}
\bibinfo{author}{\bibfnamefont{C.}~\bibnamefont{Ciofi~degli Atti}},
  \bibinfo{author}{\bibfnamefont{C.~B.} \bibnamefont{Mezzetti}},
  \bibnamefont{and} \bibinfo{author}{\bibfnamefont{H.}~\bibnamefont{Morita}},
  \bibinfo{journal}{Phys. Rev. C} \textbf{\bibinfo{volume}{95}},
  \bibinfo{pages}{044327} (\bibinfo{year}{2017}), \eprint{1701.08211}.

\bibitem[{\citenamefont{Ciofi~degli Atti and
  Morita}(2017)}]{CiofidegliAtti:2017xtx}
\bibinfo{author}{\bibfnamefont{C.}~\bibnamefont{Ciofi~degli Atti}}
  \bibnamefont{and} \bibinfo{author}{\bibfnamefont{H.}~\bibnamefont{Morita}},
  \bibinfo{journal}{Phys. Rev. C} \textbf{\bibinfo{volume}{96}},
  \bibinfo{pages}{064317} (\bibinfo{year}{2017}), \eprint{1708.05168}.

\bibitem[{\citenamefont{Alvioli et~al.}(2012)\citenamefont{Alvioli, Ciofi~degli
  Atti, Kaptari, Mezzetti, Morita, and Scopetta}}]{Alvioli:2011aa}
\bibinfo{author}{\bibfnamefont{M.}~\bibnamefont{Alvioli}},
  \bibinfo{author}{\bibfnamefont{C.}~\bibnamefont{Ciofi~degli Atti}},
  \bibinfo{author}{\bibfnamefont{L.~P.} \bibnamefont{Kaptari}},
  \bibinfo{author}{\bibfnamefont{C.~B.} \bibnamefont{Mezzetti}},
  \bibinfo{author}{\bibfnamefont{H.}~\bibnamefont{Morita}}, \bibnamefont{and}
  \bibinfo{author}{\bibfnamefont{S.}~\bibnamefont{Scopetta}},
  \bibinfo{journal}{Phys. Rev. C} \textbf{\bibinfo{volume}{85}},
  \bibinfo{pages}{021001} (\bibinfo{year}{2012}), \eprint{1112.2651}.

\bibitem[{\citenamefont{Arias~de Saavedra et~al.}(2007)\citenamefont{Arias~de
  Saavedra, Bisconti, Co', and Fabrocini}}]{AriasdeSaavedra:2007byz}
\bibinfo{author}{\bibfnamefont{F.}~\bibnamefont{Arias~de Saavedra}},
  \bibinfo{author}{\bibfnamefont{C.}~\bibnamefont{Bisconti}},
  \bibinfo{author}{\bibfnamefont{G.}~\bibnamefont{Co'}}, \bibnamefont{and}
  \bibinfo{author}{\bibfnamefont{A.}~\bibnamefont{Fabrocini}},
  \bibinfo{journal}{Phys. Rept.} \textbf{\bibinfo{volume}{450}},
  \bibinfo{pages}{1} (\bibinfo{year}{2007}), \eprint{0706.3792}.

\bibitem[{\citenamefont{Bisconti et~al.}(2007)\citenamefont{Bisconti, Arias~de
  Saavedra, and Co}}]{Bisconti:2007vu}
\bibinfo{author}{\bibfnamefont{C.}~\bibnamefont{Bisconti}},
  \bibinfo{author}{\bibfnamefont{F.}~\bibnamefont{Arias~de Saavedra}},
  \bibnamefont{and} \bibinfo{author}{\bibfnamefont{G.}~\bibnamefont{Co}},
  \bibinfo{journal}{Phys. Rev. C} \textbf{\bibinfo{volume}{75}},
  \bibinfo{pages}{054302} (\bibinfo{year}{2007}), \eprint{nucl-th/0702061}.

\bibitem[{\citenamefont{Ryckebusch et~al.}(2015)\citenamefont{Ryckebusch,
  Cosyn, and Vanhalst}}]{Ryckebusch:2014ann}
\bibinfo{author}{\bibfnamefont{J.}~\bibnamefont{Ryckebusch}},
  \bibinfo{author}{\bibfnamefont{W.}~\bibnamefont{Cosyn}}, \bibnamefont{and}
  \bibinfo{author}{\bibfnamefont{M.}~\bibnamefont{Vanhalst}},
  \bibinfo{journal}{J. Phys. G} \textbf{\bibinfo{volume}{42}},
  \bibinfo{pages}{055104} (\bibinfo{year}{2015}), \eprint{1405.3814}.

\bibitem[{\citenamefont{Wiringa et~al.}(1995)\citenamefont{Wiringa, Stoks, and
  Schiavilla}}]{Wiringa:1994wb}
\bibinfo{author}{\bibfnamefont{R.~B.} \bibnamefont{Wiringa}},
  \bibinfo{author}{\bibfnamefont{V.~G.~J.} \bibnamefont{Stoks}},
  \bibnamefont{and}
  \bibinfo{author}{\bibfnamefont{R.}~\bibnamefont{Schiavilla}},
  \bibinfo{journal}{Phys. Rev. C} \textbf{\bibinfo{volume}{51}},
  \bibinfo{pages}{38} (\bibinfo{year}{1995}), \eprint{nucl-th/9408016}.

\bibitem[{\citenamefont{Cruz-Torres et~al.}(2021)}]{Cruz-Torres:2019fum}
\bibinfo{author}{\bibfnamefont{R.}~\bibnamefont{Cruz-Torres}}
  \bibnamefont{et~al.}, \bibinfo{journal}{Nature Phys.}
  \textbf{\bibinfo{volume}{17}}, \bibinfo{pages}{306} (\bibinfo{year}{2021}),
  \eprint{1907.03658}.

\bibitem[{\citenamefont{Piarulli et~al.}(2016)\citenamefont{Piarulli, Girlanda,
  Schiavilla, Kievsky, Lovato, Marcucci, Pieper, Viviani, and
  Wiringa}}]{Piarulli:2016vel}
\bibinfo{author}{\bibfnamefont{M.}~\bibnamefont{Piarulli}},
  \bibinfo{author}{\bibfnamefont{L.}~\bibnamefont{Girlanda}},
  \bibinfo{author}{\bibfnamefont{R.}~\bibnamefont{Schiavilla}},
  \bibinfo{author}{\bibfnamefont{A.}~\bibnamefont{Kievsky}},
  \bibinfo{author}{\bibfnamefont{A.}~\bibnamefont{Lovato}},
  \bibinfo{author}{\bibfnamefont{L.~E.} \bibnamefont{Marcucci}},
  \bibinfo{author}{\bibfnamefont{S.~C.} \bibnamefont{Pieper}},
  \bibinfo{author}{\bibfnamefont{M.}~\bibnamefont{Viviani}}, \bibnamefont{and}
  \bibinfo{author}{\bibfnamefont{R.~B.} \bibnamefont{Wiringa}},
  \bibinfo{journal}{Phys. Rev. C} \textbf{\bibinfo{volume}{94}},
  \bibinfo{pages}{054007} (\bibinfo{year}{2016}), \eprint{1606.06335}.

\bibitem[{\citenamefont{Piarulli et~al.}(2015)\citenamefont{Piarulli, Girlanda,
  Schiavilla, Navarro~Pérez, Amaro, and Ruiz~Arriola}}]{Piarulli:2014bda}
\bibinfo{author}{\bibfnamefont{M.}~\bibnamefont{Piarulli}},
  \bibinfo{author}{\bibfnamefont{L.}~\bibnamefont{Girlanda}},
  \bibinfo{author}{\bibfnamefont{R.}~\bibnamefont{Schiavilla}},
  \bibinfo{author}{\bibfnamefont{R.}~\bibnamefont{Navarro~Pérez}},
  \bibinfo{author}{\bibfnamefont{J.~E.} \bibnamefont{Amaro}}, \bibnamefont{and}
  \bibinfo{author}{\bibfnamefont{E.}~\bibnamefont{Ruiz~Arriola}},
  \bibinfo{journal}{Phys. Rev. C} \textbf{\bibinfo{volume}{91}},
  \bibinfo{pages}{024003} (\bibinfo{year}{2015}), \eprint{1412.6446}.

\bibitem[{\citenamefont{Baroni et~al.}(2018)}]{Baroni:2018fdn}
\bibinfo{author}{\bibfnamefont{A.}~\bibnamefont{Baroni}} \bibnamefont{et~al.},
  \bibinfo{journal}{Phys. Rev.} \textbf{\bibinfo{volume}{C98}},
  \bibinfo{pages}{044003} (\bibinfo{year}{2018}), \eprint{1806.10245}.

\bibitem[{\citenamefont{Piarulli and Schiavilla}(2021)}]{Piarulli:2021ywl}
\bibinfo{author}{\bibfnamefont{M.}~\bibnamefont{Piarulli}} \bibnamefont{and}
  \bibinfo{author}{\bibfnamefont{R.}~\bibnamefont{Schiavilla}},
  \bibinfo{journal}{Few Body Syst.} \textbf{\bibinfo{volume}{62}},
  \bibinfo{pages}{108} (\bibinfo{year}{2021}), \eprint{2111.00675}.

\bibitem[{\citenamefont{Piarulli and Tews}(2020)}]{Piarulli:2019cqu}
\bibinfo{author}{\bibfnamefont{M.}~\bibnamefont{Piarulli}} \bibnamefont{and}
  \bibinfo{author}{\bibfnamefont{I.}~\bibnamefont{Tews}},
  \bibinfo{journal}{Front. in Phys.} \textbf{\bibinfo{volume}{7}},
  \bibinfo{pages}{245} (\bibinfo{year}{2020}), \eprint{2002.00032}.

\bibitem[{\citenamefont{Navarro~Pérez
  et~al.}(2013)\citenamefont{Navarro~Pérez, Amaro, and
  Ruiz~Arriola}}]{Perez:2013jpa}
\bibinfo{author}{\bibfnamefont{R.}~\bibnamefont{Navarro~Pérez}},
  \bibinfo{author}{\bibfnamefont{J.~E.} \bibnamefont{Amaro}}, \bibnamefont{and}
  \bibinfo{author}{\bibfnamefont{E.}~\bibnamefont{Ruiz~Arriola}},
  \bibinfo{journal}{Phys. Rev. C} \textbf{\bibinfo{volume}{88}},
  \bibinfo{pages}{064002} (\bibinfo{year}{2013}), \bibinfo{note}{[Erratum:
  Phys. Rev.C91,no.2,029901(2015)]}, \eprint{1310.2536}.

\bibitem[{\citenamefont{Navarro~Pérez
  et~al.}(2014)\citenamefont{Navarro~Pérez, Amaro, and
  Ruiz~Arriola}}]{Perez:2013oba}
\bibinfo{author}{\bibfnamefont{R.}~\bibnamefont{Navarro~Pérez}},
  \bibinfo{author}{\bibfnamefont{J.~E.} \bibnamefont{Amaro}}, \bibnamefont{and}
  \bibinfo{author}{\bibfnamefont{E.}~\bibnamefont{Ruiz~Arriola}},
  \bibinfo{journal}{Phys. Rev. C} \textbf{\bibinfo{volume}{89}},
  \bibinfo{pages}{024004} (\bibinfo{year}{2014}), \eprint{1310.6972}.

\bibitem[{\citenamefont{Navarro~Perez et~al.}(2014)\citenamefont{Navarro~Perez,
  Amaro, and Ruiz~Arriola}}]{Perez:2014yla}
\bibinfo{author}{\bibfnamefont{R.}~\bibnamefont{Navarro~Perez}},
  \bibinfo{author}{\bibfnamefont{J.~E.} \bibnamefont{Amaro}}, \bibnamefont{and}
  \bibinfo{author}{\bibfnamefont{E.}~\bibnamefont{Ruiz~Arriola}},
  \bibinfo{journal}{Phys. Rev. C} \textbf{\bibinfo{volume}{89}},
  \bibinfo{pages}{064006} (\bibinfo{year}{2014}), \eprint{1404.0314}.

\bibitem[{\citenamefont{Piarulli et~al.}(2018)}]{Piarulli:2017dwd}
\bibinfo{author}{\bibfnamefont{M.}~\bibnamefont{Piarulli}}
  \bibnamefont{et~al.}, \bibinfo{journal}{Phys. Rev. Lett.}
  \textbf{\bibinfo{volume}{120}}, \bibinfo{pages}{052503}
  (\bibinfo{year}{2018}), \eprint{1707.02883}.

\bibitem[{\citenamefont{Gazit et~al.}(2009)\citenamefont{Gazit, Quaglioni, and
  Navratil}}]{Gazit:2008ma}
\bibinfo{author}{\bibfnamefont{D.}~\bibnamefont{Gazit}},
  \bibinfo{author}{\bibfnamefont{S.}~\bibnamefont{Quaglioni}},
  \bibnamefont{and} \bibinfo{author}{\bibfnamefont{P.}~\bibnamefont{Navratil}},
  \bibinfo{journal}{Phys. Rev. Lett.} \textbf{\bibinfo{volume}{103}},
  \bibinfo{pages}{102502} (\bibinfo{year}{2009}), \eprint{0812.4444}.

\bibitem[{\citenamefont{Marcucci et~al.}(2012)\citenamefont{Marcucci, Kievsky,
  Rosati, Schiavilla, and Viviani}}]{Marcucci:2011jm}
\bibinfo{author}{\bibfnamefont{L.~E.} \bibnamefont{Marcucci}},
  \bibinfo{author}{\bibfnamefont{A.}~\bibnamefont{Kievsky}},
  \bibinfo{author}{\bibfnamefont{S.}~\bibnamefont{Rosati}},
  \bibinfo{author}{\bibfnamefont{R.}~\bibnamefont{Schiavilla}},
  \bibnamefont{and} \bibinfo{author}{\bibfnamefont{M.}~\bibnamefont{Viviani}},
  \bibinfo{journal}{Phys. Rev. Lett.} \textbf{\bibinfo{volume}{108}},
  \bibinfo{pages}{052502} (\bibinfo{year}{2012}), \bibinfo{note}{[Erratum:
  Phys. Rev. Lett.121,no.4,049901(2018)]}, \eprint{1109.5563}.

\bibitem[{\citenamefont{Schiavilla}(2017)}]{Schiavilla:2017}
\bibinfo{author}{\bibfnamefont{R.}~\bibnamefont{Schiavilla}},
  \bibinfo{journal}{unpublished}  (\bibinfo{year}{2017}).

\bibitem[{\citenamefont{Carlson et~al.}(2015)\citenamefont{Carlson, Gandolfi,
  Pederiva, Pieper, Schiavilla, Schmidt, and Wiringa}}]{Carlson:2014vla}
\bibinfo{author}{\bibfnamefont{J.}~\bibnamefont{Carlson}},
  \bibinfo{author}{\bibfnamefont{S.}~\bibnamefont{Gandolfi}},
  \bibinfo{author}{\bibfnamefont{F.}~\bibnamefont{Pederiva}},
  \bibinfo{author}{\bibfnamefont{S.~C.} \bibnamefont{Pieper}},
  \bibinfo{author}{\bibfnamefont{R.}~\bibnamefont{Schiavilla}},
  \bibinfo{author}{\bibfnamefont{K.~E.} \bibnamefont{Schmidt}},
  \bibnamefont{and} \bibinfo{author}{\bibfnamefont{R.~B.}
  \bibnamefont{Wiringa}}, \bibinfo{journal}{Rev. Mod. Phys.}
  \textbf{\bibinfo{volume}{87}}, \bibinfo{pages}{1067} (\bibinfo{year}{2015}),
  \eprint{1412.3081}.

\bibitem[{\citenamefont{Gandolfi et~al.}(2020)\citenamefont{Gandolfi,
  Lonardoni, Lovato, and Piarulli}}]{Gandolfi:2020pbj}
\bibinfo{author}{\bibfnamefont{S.}~\bibnamefont{Gandolfi}},
  \bibinfo{author}{\bibfnamefont{D.}~\bibnamefont{Lonardoni}},
  \bibinfo{author}{\bibfnamefont{A.}~\bibnamefont{Lovato}}, \bibnamefont{and}
  \bibinfo{author}{\bibfnamefont{M.}~\bibnamefont{Piarulli}},
  \bibinfo{journal}{Front. in Phys.} \textbf{\bibinfo{volume}{8}},
  \bibinfo{pages}{117} (\bibinfo{year}{2020}), \eprint{2001.01374}.

\bibitem[{\citenamefont{King et~al.}(2020{\natexlab{a}})\citenamefont{King,
  Andreoli, Pastore, and Piarulli}}]{King:2020pza}
\bibinfo{author}{\bibfnamefont{G.~B.} \bibnamefont{King}},
  \bibinfo{author}{\bibfnamefont{L.}~\bibnamefont{Andreoli}},
  \bibinfo{author}{\bibfnamefont{S.}~\bibnamefont{Pastore}}, \bibnamefont{and}
  \bibinfo{author}{\bibfnamefont{M.}~\bibnamefont{Piarulli}},
  \bibinfo{journal}{Front. in Phys.} \textbf{\bibinfo{volume}{8}},
  \bibinfo{pages}{363} (\bibinfo{year}{2020}{\natexlab{a}}).

\bibitem[{\citenamefont{King et~al.}(2020{\natexlab{b}})\citenamefont{King,
  Andreoli, Pastore, Piarulli, Schiavilla, Wiringa, Carlson, and
  Gandolfi}}]{King:2020wmp}
\bibinfo{author}{\bibfnamefont{G.~B.} \bibnamefont{King}},
  \bibinfo{author}{\bibfnamefont{L.}~\bibnamefont{Andreoli}},
  \bibinfo{author}{\bibfnamefont{S.}~\bibnamefont{Pastore}},
  \bibinfo{author}{\bibfnamefont{M.}~\bibnamefont{Piarulli}},
  \bibinfo{author}{\bibfnamefont{R.}~\bibnamefont{Schiavilla}},
  \bibinfo{author}{\bibfnamefont{R.~B.} \bibnamefont{Wiringa}},
  \bibinfo{author}{\bibfnamefont{J.}~\bibnamefont{Carlson}}, \bibnamefont{and}
  \bibinfo{author}{\bibfnamefont{S.}~\bibnamefont{Gandolfi}},
  \bibinfo{journal}{Phys. Rev. C} \textbf{\bibinfo{volume}{102}},
  \bibinfo{pages}{025501} (\bibinfo{year}{2020}{\natexlab{b}}),
  \eprint{2004.05263}.

\bibitem[{\citenamefont{Cirigliano et~al.}(2019)\citenamefont{Cirigliano,
  Dekens, De~Vries, Graesser, Mereghetti, Pastore, Piarulli, Van~Kolck, and
  Wiringa}}]{Cirigliano:2019vdj}
\bibinfo{author}{\bibfnamefont{V.}~\bibnamefont{Cirigliano}},
  \bibinfo{author}{\bibfnamefont{W.}~\bibnamefont{Dekens}},
  \bibinfo{author}{\bibfnamefont{J.}~\bibnamefont{De~Vries}},
  \bibinfo{author}{\bibfnamefont{M.~L.} \bibnamefont{Graesser}},
  \bibinfo{author}{\bibfnamefont{E.}~\bibnamefont{Mereghetti}},
  \bibinfo{author}{\bibfnamefont{S.}~\bibnamefont{Pastore}},
  \bibinfo{author}{\bibfnamefont{M.}~\bibnamefont{Piarulli}},
  \bibinfo{author}{\bibfnamefont{U.}~\bibnamefont{Van~Kolck}},
  \bibnamefont{and} \bibinfo{author}{\bibfnamefont{R.~B.}
  \bibnamefont{Wiringa}}, \bibinfo{journal}{Phys. Rev. C}
  \textbf{\bibinfo{volume}{100}}, \bibinfo{pages}{055504}
  (\bibinfo{year}{2019}), \eprint{1907.11254}.

\bibitem[{\citenamefont{Cirigliano et~al.}(2018)\citenamefont{Cirigliano,
  Dekens, De~Vries, Graesser, Mereghetti, Pastore, and
  Van~Kolck}}]{Cirigliano:2018hja}
\bibinfo{author}{\bibfnamefont{V.}~\bibnamefont{Cirigliano}},
  \bibinfo{author}{\bibfnamefont{W.}~\bibnamefont{Dekens}},
  \bibinfo{author}{\bibfnamefont{J.}~\bibnamefont{De~Vries}},
  \bibinfo{author}{\bibfnamefont{M.~L.} \bibnamefont{Graesser}},
  \bibinfo{author}{\bibfnamefont{E.}~\bibnamefont{Mereghetti}},
  \bibinfo{author}{\bibfnamefont{S.}~\bibnamefont{Pastore}}, \bibnamefont{and}
  \bibinfo{author}{\bibfnamefont{U.}~\bibnamefont{Van~Kolck}},
  \bibinfo{journal}{Phys. Rev. Lett.} \textbf{\bibinfo{volume}{120}},
  \bibinfo{pages}{202001} (\bibinfo{year}{2018}), \eprint{1802.10097}.

\bibitem[{\citenamefont{King et~al.}(2022{\natexlab{a}})\citenamefont{King,
  Baroni, Cirigliano, Gandolfi, Hayen, Mereghetti, Pastore, and
  Piarulli}}]{King:2022zkz}
\bibinfo{author}{\bibfnamefont{G.~B.} \bibnamefont{King}},
  \bibinfo{author}{\bibfnamefont{A.}~\bibnamefont{Baroni}},
  \bibinfo{author}{\bibfnamefont{V.}~\bibnamefont{Cirigliano}},
  \bibinfo{author}{\bibfnamefont{S.}~\bibnamefont{Gandolfi}},
  \bibinfo{author}{\bibfnamefont{L.}~\bibnamefont{Hayen}},
  \bibinfo{author}{\bibfnamefont{E.}~\bibnamefont{Mereghetti}},
  \bibinfo{author}{\bibfnamefont{S.}~\bibnamefont{Pastore}}, \bibnamefont{and}
  \bibinfo{author}{\bibfnamefont{M.}~\bibnamefont{Piarulli}}
  (\bibinfo{year}{2022}{\natexlab{a}}), \eprint{2207.11179}.

\bibitem[{\citenamefont{King et~al.}(2022{\natexlab{b}})\citenamefont{King,
  Pastore, Piarulli, and Schiavilla}}]{King:2021jdb}
\bibinfo{author}{\bibfnamefont{G.~B.} \bibnamefont{King}},
  \bibinfo{author}{\bibfnamefont{S.}~\bibnamefont{Pastore}},
  \bibinfo{author}{\bibfnamefont{M.}~\bibnamefont{Piarulli}}, \bibnamefont{and}
  \bibinfo{author}{\bibfnamefont{R.}~\bibnamefont{Schiavilla}},
  \bibinfo{journal}{Phys. Rev. C} \textbf{\bibinfo{volume}{105}},
  \bibinfo{pages}{L042501} (\bibinfo{year}{2022}{\natexlab{b}}),
  \eprint{2111.11360}.

\bibitem[{\citenamefont{Piarulli et~al.}(2020)\citenamefont{Piarulli, Bombaci,
  Logoteta, Lovato, and Wiringa}}]{Piarulli:2019pfq}
\bibinfo{author}{\bibfnamefont{M.}~\bibnamefont{Piarulli}},
  \bibinfo{author}{\bibfnamefont{I.}~\bibnamefont{Bombaci}},
  \bibinfo{author}{\bibfnamefont{D.}~\bibnamefont{Logoteta}},
  \bibinfo{author}{\bibfnamefont{A.}~\bibnamefont{Lovato}}, \bibnamefont{and}
  \bibinfo{author}{\bibfnamefont{R.~B.} \bibnamefont{Wiringa}},
  \bibinfo{journal}{Phys. Rev. C} \textbf{\bibinfo{volume}{101}},
  \bibinfo{pages}{045801} (\bibinfo{year}{2020}), \eprint{1908.04426}.

\bibitem[{\citenamefont{Lovato et~al.}(2022)\citenamefont{Lovato, Bombaci,
  Logoteta, Piarulli, and Wiringa}}]{Lovato:2022apd}
\bibinfo{author}{\bibfnamefont{A.}~\bibnamefont{Lovato}},
  \bibinfo{author}{\bibfnamefont{I.}~\bibnamefont{Bombaci}},
  \bibinfo{author}{\bibfnamefont{D.}~\bibnamefont{Logoteta}},
  \bibinfo{author}{\bibfnamefont{M.}~\bibnamefont{Piarulli}}, \bibnamefont{and}
  \bibinfo{author}{\bibfnamefont{R.~B.} \bibnamefont{Wiringa}},
  \bibinfo{journal}{Phys. Rev. C} \textbf{\bibinfo{volume}{105}}
  (\bibinfo{year}{2022}), \eprint{2202.10293}.

\bibitem[{\citenamefont{Pudliner et~al.}(1997)\citenamefont{Pudliner,
  Pandharipande, Carlson, Pieper, and Wiringa}}]{Pudliner:1997ck}
\bibinfo{author}{\bibfnamefont{B.~S.} \bibnamefont{Pudliner}},
  \bibinfo{author}{\bibfnamefont{V.~R.} \bibnamefont{Pandharipande}},
  \bibinfo{author}{\bibfnamefont{J.}~\bibnamefont{Carlson}},
  \bibinfo{author}{\bibfnamefont{S.~C.} \bibnamefont{Pieper}},
  \bibnamefont{and} \bibinfo{author}{\bibfnamefont{R.~B.}
  \bibnamefont{Wiringa}}, \bibinfo{journal}{Phys. Rev. C}
  \textbf{\bibinfo{volume}{56}}, \bibinfo{pages}{1720} (\bibinfo{year}{1997}),
  \eprint{nucl-th/9705009}.

\bibitem[{\citenamefont{Pastore et~al.}(2020)\citenamefont{Pastore, Carlson,
  Gandolfi, Schiavilla, and Wiringa}}]{Pastore:2019urn}
\bibinfo{author}{\bibfnamefont{S.}~\bibnamefont{Pastore}},
  \bibinfo{author}{\bibfnamefont{J.}~\bibnamefont{Carlson}},
  \bibinfo{author}{\bibfnamefont{S.}~\bibnamefont{Gandolfi}},
  \bibinfo{author}{\bibfnamefont{R.}~\bibnamefont{Schiavilla}},
  \bibnamefont{and} \bibinfo{author}{\bibfnamefont{R.~B.}
  \bibnamefont{Wiringa}}, \bibinfo{journal}{Phys. Rev. C}
  \textbf{\bibinfo{volume}{101}}, \bibinfo{pages}{044612}
  (\bibinfo{year}{2020}), \eprint{1909.06400}.

\bibitem[{\citenamefont{Barrow et~al.}(2021)\citenamefont{Barrow, Gardiner,
  Pastore, Betancourt, and Carlson}}]{Barrow:2020mfy}
\bibinfo{author}{\bibfnamefont{J.~L.} \bibnamefont{Barrow}},
  \bibinfo{author}{\bibfnamefont{S.}~\bibnamefont{Gardiner}},
  \bibinfo{author}{\bibfnamefont{S.}~\bibnamefont{Pastore}},
  \bibinfo{author}{\bibfnamefont{M.}~\bibnamefont{Betancourt}},
  \bibnamefont{and} \bibinfo{author}{\bibfnamefont{J.}~\bibnamefont{Carlson}},
  \bibinfo{journal}{Phys. Rev. D} \textbf{\bibinfo{volume}{103}},
  \bibinfo{pages}{052001} (\bibinfo{year}{2021}), \eprint{2010.04154}.

\bibitem[{\citenamefont{Andreoli et~al.}(2022)\citenamefont{Andreoli, Carlson,
  Lovato, Pastore, Rocco, and Wiringa}}]{Andreoli:2021cxo}
\bibinfo{author}{\bibfnamefont{L.}~\bibnamefont{Andreoli}},
  \bibinfo{author}{\bibfnamefont{J.}~\bibnamefont{Carlson}},
  \bibinfo{author}{\bibfnamefont{A.}~\bibnamefont{Lovato}},
  \bibinfo{author}{\bibfnamefont{S.}~\bibnamefont{Pastore}},
  \bibinfo{author}{\bibfnamefont{N.}~\bibnamefont{Rocco}}, \bibnamefont{and}
  \bibinfo{author}{\bibfnamefont{R.~B.} \bibnamefont{Wiringa}},
  \bibinfo{journal}{Phys. Rev. C} \textbf{\bibinfo{volume}{105}},
  \bibinfo{pages}{014002} (\bibinfo{year}{2022}), \eprint{2108.10824}.

\end{thebibliography}
\end{document}